\documentclass[journal,twoside]{IEEEtran}
\usepackage{graphicx,hyperref,color,rotating,amsmath,amssymb}
\newcommand{\ie}{{\it i.e.\/} }

\newcommand{\cf}{{\it cf.\/} }

\begin{document}
\title{Mechanical Measurements of the ALMA Prototype Antennas}

\author{A.~Greve 
        and J.~G.~Mangum
\thanks{Manuscript received May 15, 2006; revised August 17, 2007.}
\thanks{The performance results presented in this publication were
  part of a comprehensive technical evaluation process used to
  evaluate the ALMA prototype antennas which concluded in April 2005.}}

\markboth{IEEE Transactions on Antennas and Propagation Magazine,~Vol.~50,
  No.~2,~April~2008}{Greve \& Mangum: Mechanical Measurements of
  the ALMA Prototype Antennas}
%

\pubid{1045--9243/00\$00.00~\copyright~2008 IEEE}


\maketitle

\begin{abstract}

The specifications of the Atacama Large Millimeter Array (ALMA) have placed 
stringent requirements on the mechanical performance of its antennas. As part 
of the evaluation process of the VertexRSI and Alcatel EIE Consortium (AEC) 
ALMA prototype antennas, measurements of the path length, thermal, and azimuth
bearing performance were made under a variety of weather conditions and 
observing modes. The results of mechanical measurements, reported here, are 
compared to the antenna specifications.

\end{abstract}


\IEEEpeerreviewmaketitle
%

\section{Introduction}

\PARstart{T}{he} Atacama Large Millimeter Array (ALMA) for astronomical
observations at 
millimeter and sub--millimeter wavelengths (up to the Terra--Hertz
region) needs antennas of high mechanical precision and of
understandable and predictable behaviour.  This 
behaviour must be established for structural deformations due to gravity, 
temperature changes, and wind loads. This means, in particular, that a high 
reflector surface precision, pointing and phase stability must be maintained
under all motions of tracking and mapping. We present a summary of tests of 
the mechanical and thermal behaviour of the 12m diameter VertexRSI and AEC 
ALMA prototype antennas, built at the VLA site (2000m altitude), New Mexico,
USA. The tests were made at several intervals between March 2003 and April 
2004, and concentrated primarily on the verification of the antenna 
specifications, of path length variations and parameters which 
influence the pointing. The data were also analyzed to understand the general 
behaviour of the antennas. In the investigation we have paid attention
to the fact 
that variations in the behaviour of the antennas may be predictable or 
sporadic. We believe that repeatable and/or predictable variations can to a 
large extent be considered in the pointing model, or any other correction 
device. The antennas were tested in stationary position, under sidereal 
tracking, On--the--Fly (OTF) mapping, and in Fast--Switching mode (FSW). A 
large amount of data was collected during commissioning and thus refer to 
all types of tracking, OTF, FSW, and unintended 'shaking'. 

A more extended summary of these test results was reported by the Antenna 
Evaluation Group to the National Radio Astronomy Observatory (NRAO)
and European Southern Observatory (ESO) (which forms the ALMA
partnership) for selection of the ALMA production antenna(s).  An
overview of these performance results was presented in
\cite{Mangum2006}.  The current paper presents a more detailed
analysis of the mechanical performance measurements made of the ALMA
prototype antennas.

\section{Structural Design Characteristics}

\subsection{VertexRSI Prototype}

\begin{figure}
  \resizebox{\hsize}{!}{
  \includegraphics[scale=1.00,angle=0]{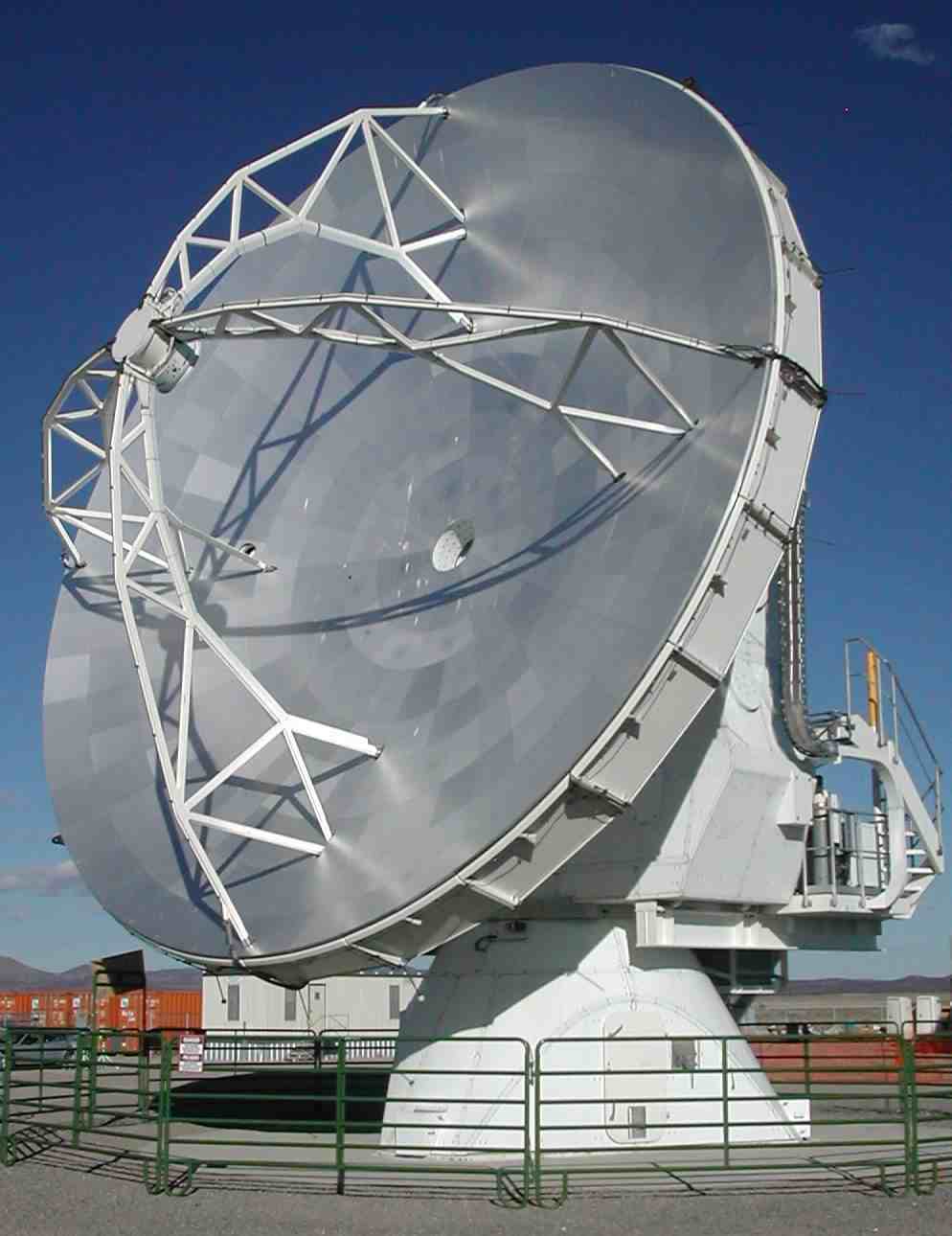}
  \includegraphics[scale=0.336,angle=0]{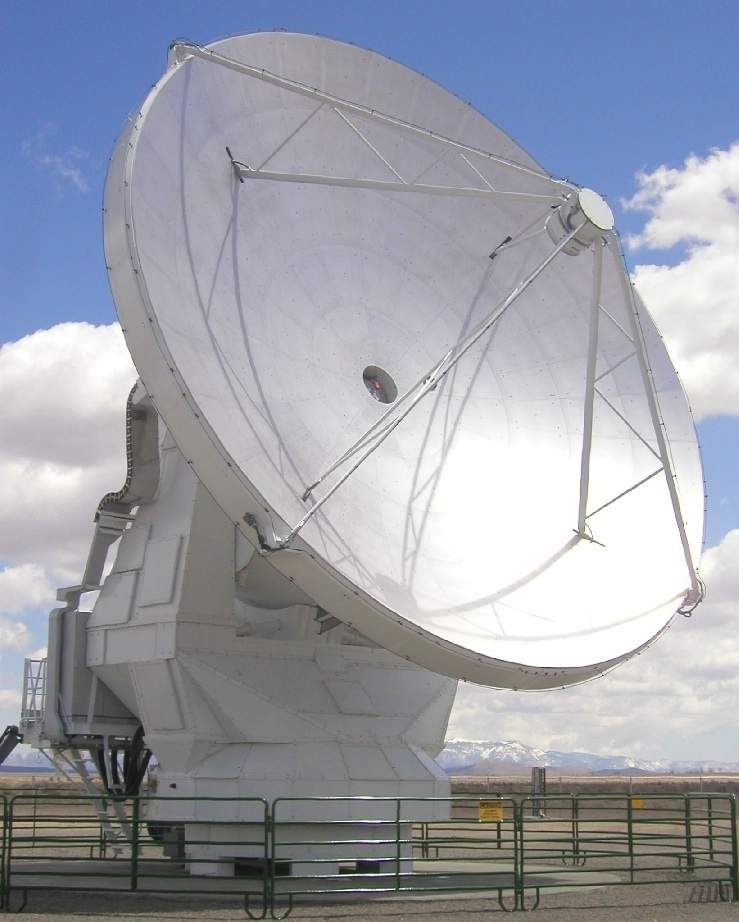}}
  \caption{The VertexRSI (left) and AEC (right) ALMA prototype
    antennas (from \cite{Mangum2006}).}
  \label{fig:antennapics}
\end{figure}

The VertexRSI antenna (Figure~\ref{fig:antennapics}) uses four
materials, \ie steel, an Invar cone, 
low thermal expansion CFRP (carbon fibre reinforced plastic), and 
CFRP--covered aluminum honeycomb plates. The 
antenna consists of a triangular pedestal; the azimuth (AZ) bearing;
the traverse and  
the fork arms (the fork); the receiver cabin and Invar cone; the backup 
structure (BUS); and the quadripod and subreflector. The BUS is made of low 
thermal--expansion CFRP--plated aluminum honeycomb; the quadripod is made of 
CFRP; the BUS support cone (on the focus cabin) is made of Invar. The
steel focus cabin is thermally controlled and stabilized (Freon
system). The pedestal and the fork 
are made of steel. The steel parts are covered with thermal insulation (foam).
The antenna is painted white. The prototype subreflector is made of
aluminum, attached  
by a 50\,cm long aluminum tube to the quadripod. The antenna is equipped with 
gear drives.  A modified copy of the ALMA VertexRSI antenna, with
special Nasmyth cabins, is the Atacama Pathfinder EXperiment (APEX)
telescope located on the Chajnantor site and in operation since
2005. A desription of this telescope and its behaviour and results at
460 and 810 GHz are published by \cite{Guesten2006}. 

\pubidadjcol 

The VertexRSI antenna contains also a metrology system, consisting 
of two tiltmeters installed in the pedestal; one tiltmeter above the AZ 
bearing; and one tiltmeter at each elevation bearing. An independent CFRP 
reference frame is mounted inside the fork. This reference structure supports,
at each elevation bearing, two linear displacement sensors, of which the 
output difference is expected to be a measure of the tilt of the antenna in 
the direction perpendicular to the elevation axis. We have analyzed the 
tilmeter readings; however, we have not tested the functionality of the 
metrology system.

In order to obtain data on the temperature distribution within the VertexRSI 
antenna, 89 temperature sensors, distributed throughout the pedestal, the 
fork, the BUS, and the quadripod and subreflector, were installed at 
customer's request. The sensors do not form a part of the metrology system. 
It is found that the measurements are useful for the understanding, or even 
prediction, of the thermal state of the antenna, and for prediction of 
temperature induced path length variations of the steel parts.

\subsection{AEC Prototype}

The AEC antenna (Figure~\ref{fig:antennapics}) uses two materials,
i.e. steel and CFRP. The AEC antenna 
consists of a cylindrical double--walled steel base (the external and 
internal pedestal), the AZ bearing, the traverse and the fork arms (the fork),
the receiver cabin, the backup structure (BUS), the quadripod and the 
subreflector. The receiver cabin, the BUS and the quadripod are made of low 
thermal--expansion CFRP plates. The other parts are made of steel. The steel 
parts are covered with thermal insulation (foam). The antenna is painted 
white. The prototype subreflector is made of aluminum, attached by a
50\,cm long aluminum tube to the quadripod. The antenna is equipped
with linear drives.

The AEC antenna contained initially a metrology system consisting of
Automated Precision Incorporated 5\,D devices (see \S\ref{instr}),
installed in both fork arms.  
The metrology system was not put into operation, and has not been tested. The 
AEC antenna contains 101 temperature sensors, distributed throughout the 
pedestal and the fork. These sensors were initially part of the metrology 
system; they have not been tested for this purpose.

A picture of the AEC prototype antenna can be found in
\cite{Mangum2006}.

\section{Scope of the Tests}

The aim of the tests was the verification of mechanical
specifications, laid down in the ``ALMA Project Statement of
Work''. This paper presents measurements, and their analysis, of

\begin{description}
\item[--] Path Length Variations,
\item[--] Thermal Behaviour,
\item[--] AZ Bearing Precision.
\end{description}

\noindent{These} tests were performed under a variety of motions of
the antenna (stationary, slewing, tracking, OTF, and FSW observations)
and meteorological conditions (stable conditions, ambient air
temperature variations,
variations of wind speed and angle of attack of the wind, different
solar illumination).
While the conditions of telescope motion were, to a large extent, 
controllable, the meteorological conditions obviously were not. This concerns 
in particular the wind conditions. 

Measurements, made during the AEG test time, of gravity induced deformations 
of the VertexRSI antenna BUS were not successfull since the supplied laser 
mount inside the reflector vertex hole was unstable against tilt. However, on 
the VertexRSI antenna the change of the subreflector distance as function of 
elevation has been determined since these measurements are not affected by 
a tilt of the mount. Gravity induced deformations were measured on the AEC 
antenna which had a stable mount inside the reflector vertex hole, used 
earlier for laser tracker surface measurements. On this antenna the change of 
the subreflector distance as function of elevation has been determined, as 
well as gravity deformations of the rim of the BUS.

\section{Employed Instrumentation}
\label{instr}

The instrumentation employed for the tests, listed in
Table~\ref{tab:instr}, consisted of:

\begin{description}
\item[--] A Laser--Diode Quadrant Detector (QD) [produced by
  FixturLaser (Sweden)]; able to measure deviations (x,y) 
perpendicular to the laser beam; 
\item[--] API 5D Measurement Device (API) [produced by Automated Precision 
Incorporated (USA)], able to measure relative changes in distance
  ($\Delta$\,z, laser interferometer), deviations (x,y) perpendicular
  to the laser beam, and perpendicular tilts of the target;
\item[--] PT\,100 Temperature Sensors [manufactured by TC Group (USA)]; of
  0.1 to 0.3$^{\rm o}$\,C precision;
\item[--] Tiltmeters [manufactured by Applied Geomechanics (USA)], of
  0.1\,arcsec precision.
\end{description}


\begin{table} 
\centering
\caption{Application of Instrumentation}
\begin{tabular}{|lll|}
\hline
Instrument & Application & Antenna \\
\hline
{\bf QD}  & Deformation Measurements:   & \\
          & Gravity Deformation BUS    & AEC \\
\hline
{\bf API} & Path--Lengths: &   \\
          & Pedestal (L$_{1}$) & VertexRSI \\
          & Fork Arm (L$_{2}$) & VertexRSI, AEC \\
          & Vertex to Subreflector (L$_{3}$) & VertexRSI, AEC \\
\hline
{\bf T--Sensors} & Temperature Behaviour: &    \\
                 & Pedestal  & VertexRSI, AEC \\
                 & Traverse  & VertexRSI, AEC \\
                 & Fork Arms & VertexRSI, AEC \\
                 & Receiver Cabin & VertexRSI \\
                 & Invar Cone  & VertexRSI\\
                 & BUS         & VertexRSI \\
                 & Quadripod $\&$ Subreflector  & VertexRSI \\
\hline
{\bf Tiltmeter}     & Tilt Measurements: &   \\
                    & AZ Axis & VertexRSI, AEC \\
\hline
\end{tabular}
\label{tab:instr}
\end{table}

\section{Path Length Measurements}

The path length specification for the ALMA prototype antenna states
that \textit{the antenna should have a non--repeatable/repeatable path
  length variation $\leq 15/20 \mu$m, respectively, within the time
  between expected consecutive calibrations of the interferometer
  array (15--30 minutes)}.

The API5D and QD measure straightness variations of a laser beam
perpendicular to the line--of--sight. The measurements are mainly
affected by atmospheric turbulence, of approximately 0.5\,arcsec (rms)
in both directions. The accuracy of open--air or enclosed measurements
over typical distances of the antenna size is $\sim$\,5\,$\mu$m. 

\begin{figure}
\resizebox{\hsize}{!}{
\includegraphics{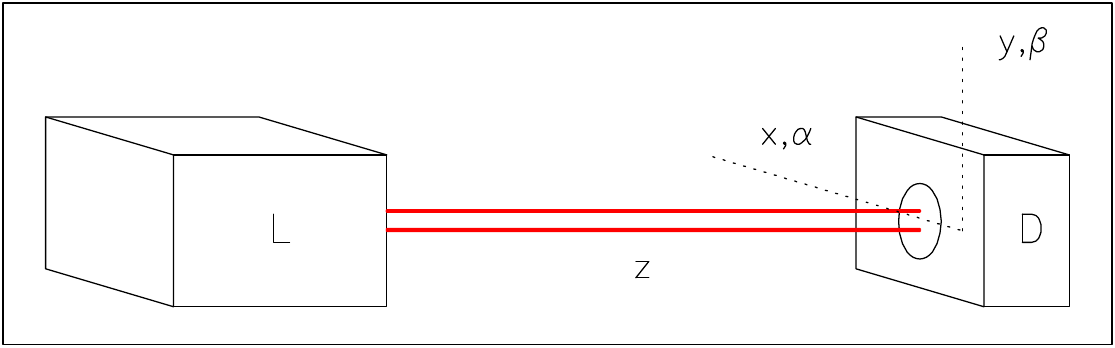}}
\caption{Schematic of the API 5\,D measurement equipment. The instrument 
measures the variation of the distance z, the position shift x,y, and the 
rotation $\alpha$,\,$\beta$ around the axes x,y. L = laser beam emitter and 
fringe counter, D = detector (target).}
\label{fig:api-schematic}
\end{figure}

\begin{figure}
\resizebox{\hsize}{!}{
\includegraphics{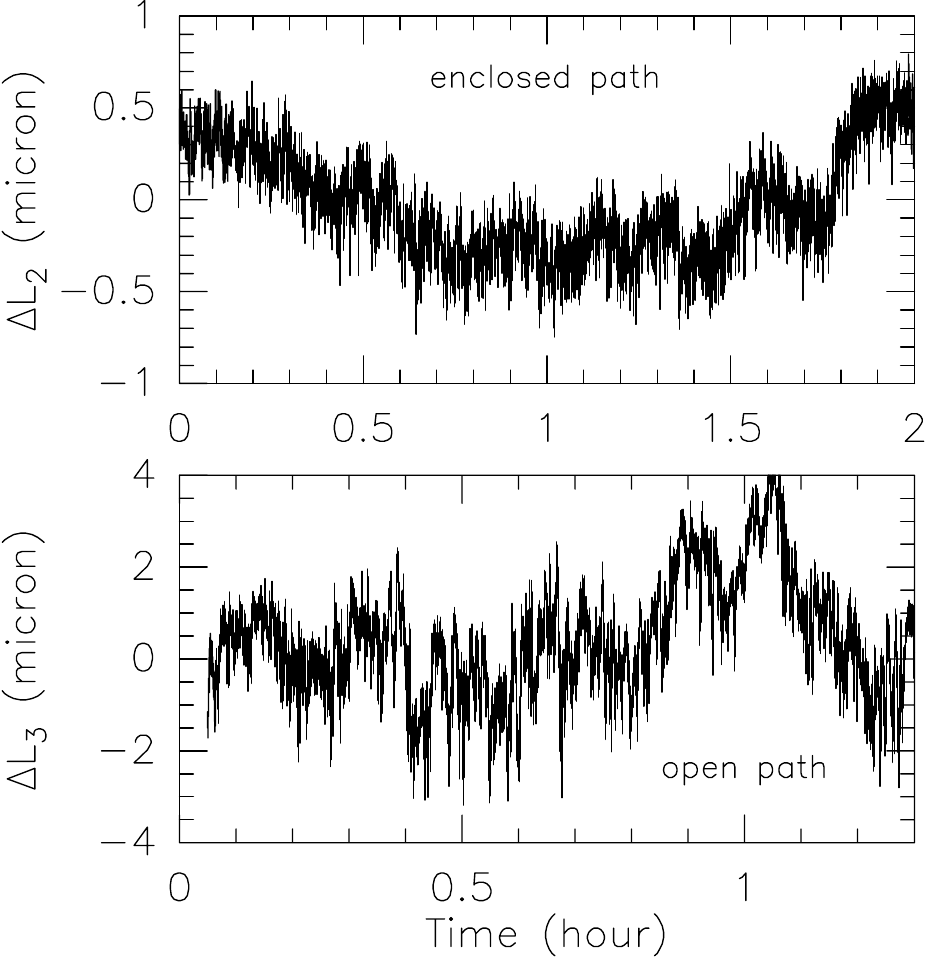}}
\caption{API measurement of path length changes ($\Delta$z) inside 
the fork arm (enclosed path) and toward the subreflector (open path).}
\label{fig:api-accuracy-z}
\end{figure}

The path lengths were measured with the API Laser Interferometer.  As
illustrated in Fig.~\ref{fig:api-schematic}, as one particular option
the API measures path length variation in the 
line--of--sight by laser interferometry. The accuracy of the measurements 
($\Delta$\,z) is better than 1\,$\mu$m for an enclosed path, and 2 to 
3\,$\mu$m micron for an open--air path, as illustrated in
Fig.~\ref{fig:api-accuracy-z}.


\begin{figure}
\resizebox{\hsize}{!}{
\includegraphics{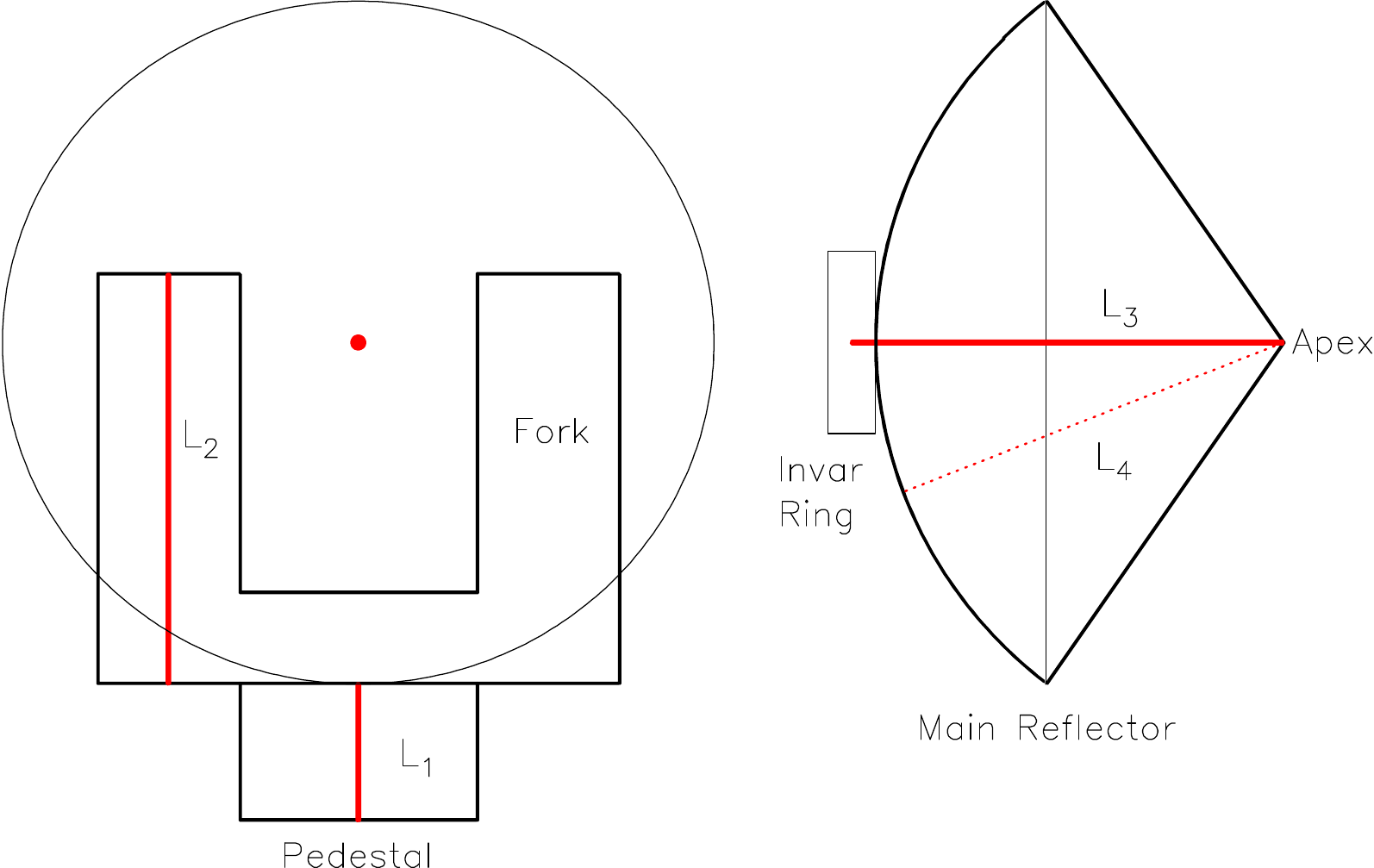}}
\caption{Illustration of the Path Length measurements L$_1$, L$_2$,
  L$_3$, and L$_4$ (from \cite{Mangum2006}).}
\label{fig:pathlength-layout}
\end{figure}

With the API
instrument it is not possible to measure the full path length variation 
between, for instance, the receiver and the ground (foundation), thus 
evaluating the integrated behaviour of the antenna as it affects the phase
of the interferometer array. The path length variations of several individual 
structural sections were measured instead, with the understanding that the  
full path length variation is approximately the sum of the components (see 
Tables \ref{tab:plsummary}). As shown in 
Fig.~\ref{fig:pathlength-layout}, these measurements contain

\begin{description}
\item[L1:] path length inside the pedestal =  distance between the foundation 
 (ground) and the lower face of the azimuth\,(AZ)--bearing; it was not possible
 to measure this path length on the AEC antenna;
\item[L2:] path length inside (one) fork arm (left,\,right) = distance between
 the lower part of the traverse and the upper part of the left (right) fork 
 arm, below the elevation (EL)-- bearing; 
\item[L3:] path length along the radio axis = distance between the upper part 
 of the Invar cone (VertexRSI), or the vertex hole (AEC, laser tracker mount) 
 of the main reflector, and the apex of the subreflector; (in the measurements
 the dilatation of the aluminum subreflector support tube (50\,cm length) has 
 been eliminated, as far as possible; see \S\ref{pathtempl3}); 
\item[L4:] path length of the reflector = distance between the Invar 
 cone/vertex hole and a point of the reflector surface, measured via the 
 subreflector. On both antennas this path length measurement was tested but 
 not routinely performed, mainly because of difficult installation and
 alignment, and time 
 limitations. Since, in essence, this path length component is the CFRP part 
 of the antennas, at least a temperature induced path length variation is
 expected to be small.
\end{description}

\subsection{Summary of Path Length Variations}
\label{pathsummary}

\begin{figure}
\resizebox{\hsize}{!}{
\includegraphics{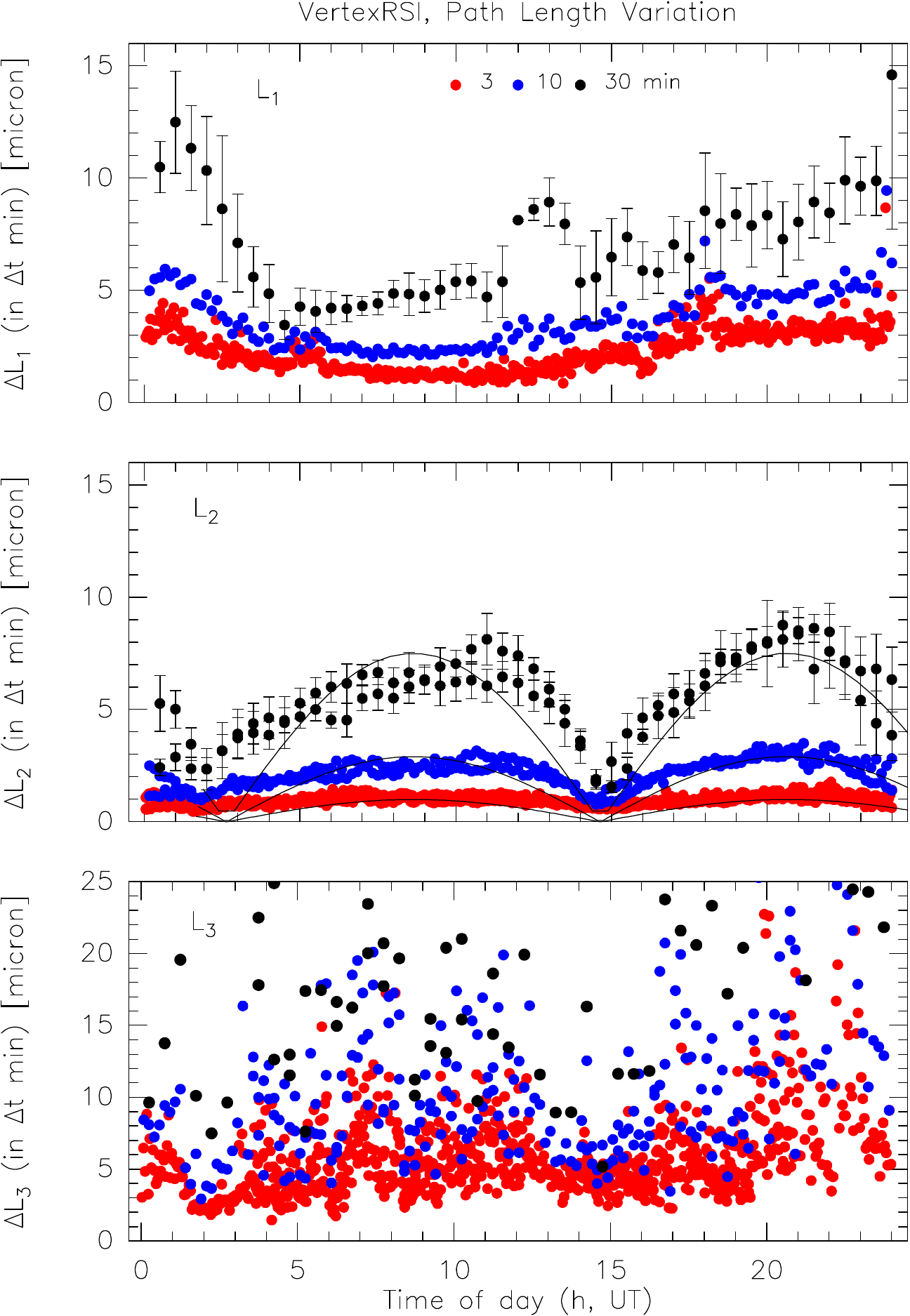}}
\caption{VertexRSI antenna. Daily variation of path length L1: pedestal, of 
 path length L2: fork arm, and of path length L3: Invar cone to subreflector, 
 as function of the time of the day (UT), and within time intervals of
 3, 10, and 30 minutes duration. The lines in the $\Delta L_2$ panel
 are explained in \S\ref{pathtempl3} (from \cite{Mangum2006}).} 
\label{fig:vertex-pl-summary}
\end{figure}

\begin{figure}
\resizebox{\hsize}{!}{
\centering
\includegraphics{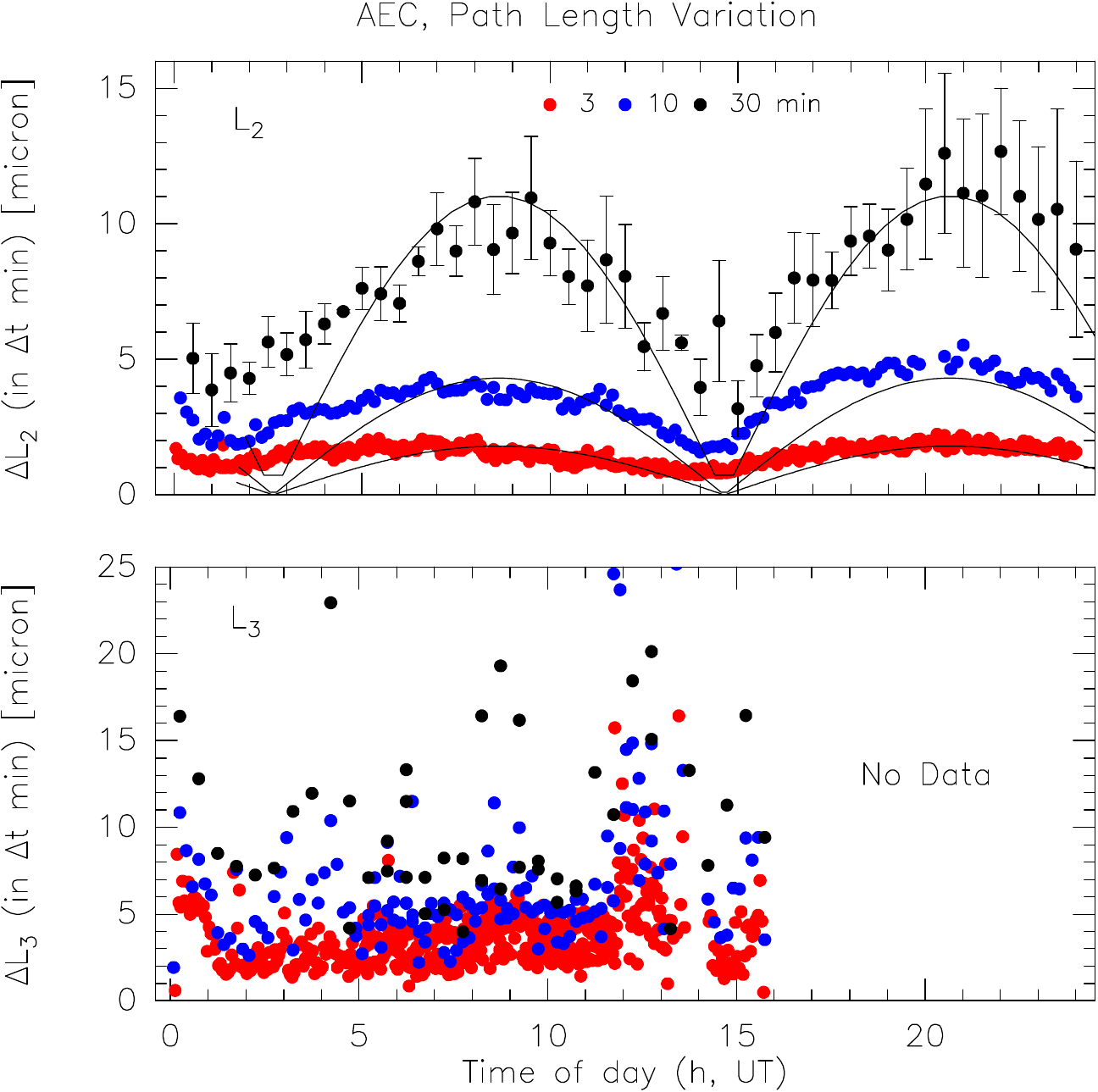}}
\caption{AEC antenna. Daily variation of pathlength L2: fork arm, and of 
path--length L3: vertex platform to subreflector, as function of the time of 
the day (UT), and within intervals of 3, 10, and 30 minutes 
duration. The lines in the $\Delta L_2$ panel are explained in
\S\ref{pathtempl3} (from \cite{Mangum2006}).} 
\label{fig:aec-pl-summary}
\end{figure}

From simple geometrical and material arguments (thermal expansion 
coefficients), and the fact that wind forces act on short time scales, it was 
evident from the beginning that the path length changes are, primarily,
due to thermal dilatation of the antenna components, induced by variation of
the ambient air temperature and solar radiation, buffered by the surface 
finish (paint) and thermal insulation. In the analysis of path length 
variations we have therefore selected time intervals ($\Delta$\,t) of
3 minutes, 
comparable to the time scale of wind, of 10 minutes, comparable to FSW and 
OTF modes of observation, and of 30 minutes, comparable to the time between 
upgrades of the pointing and interferometer phases. The corresponding path 
length measurements of the VertexRSI and the AEC antenna, as function of time 
of the day (local time = UT -- 7\,h), are shown in
Fig.~\ref{fig:vertex-pl-summary} and \ref{fig:aec-pl-summary}.  The
results are summarized in Table \ref{tab:plsummary}. The data shown in
Fig.~\ref{fig:vertex-pl-summary} and \ref{fig:aec-pl-summary} were
obtained during 
long time periods (see Table~\ref{tab:plsummary}), and cover a large variety 
of antenna motions. For path lengths which could not be measured an estimated
value is entered in the tables.

\begin{table}
\centering
\caption{Path Length Variations}
\begin{tabular}{|l|l|l|}
\hline
Path & \multicolumn{1}{c|}{VertexRSI$^a$} & \multicolumn{1}{c|}{AEC$^a$} \\
     & ($\mu$m,$\mu$m,$\mu$m,hr) & ($\mu$m,$\mu$m,$\mu$m,hr) \\
\hline
L$_{1}$\,: pedestal & 3/6/10/250 &  $\sim$\,5/$\sim$\,5/$\sim$\,5/est \\
L$_{2}$\,: fork arm & 1.5/3/8/360 &  3/4/10/360 \\
L$_{3}$\,: quadripod & 5/5/5/25  & 4/5/$\sim$\,10/25 \\
L$_{4}$\,: reflector & $\sim$\,5/$\sim$\,7/$\sim$\,9/est &
$\sim$\,5/$\sim$\,5/$\sim$\,5/est \\
\hline
$\Delta$\,z & 15/21/32/635 & 15/18/30/385 \\
\hline
\multicolumn{3}{l}{$^a$~Each measurement listed as averages over 3,
  10, and} \\
\multicolumn{3}{l}{~~~30 minute durations, along with the total measurement} \\
\multicolumn{3}{l}{~~~time (or estimate), for each path.} \\
\end{tabular}
\label{tab:plsummary}
\end{table}

\subsection{Path Length Variations Influenced by Temperature and Wind}
\label{pathtempwind}

\begin{figure}
\resizebox{\hsize}{!}{
\includegraphics[angle=-90]{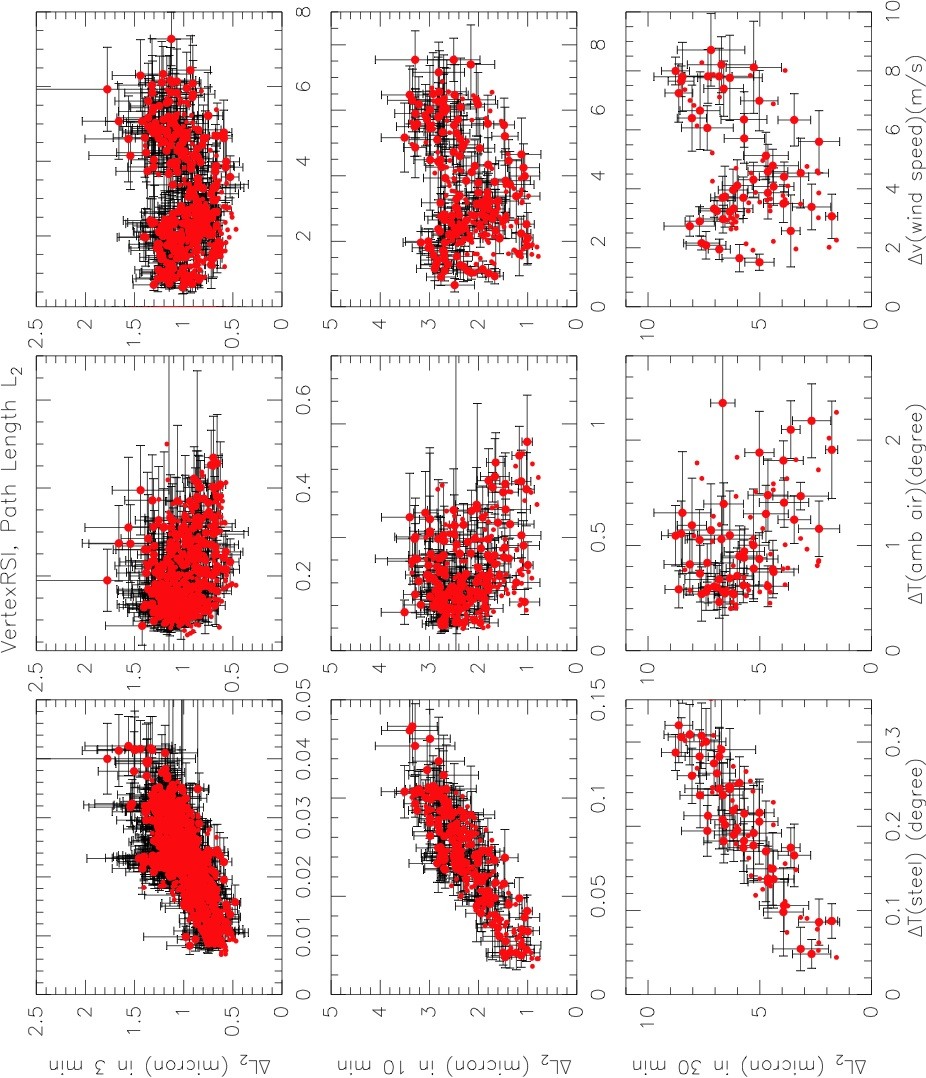}}
\caption{VertexRSI antenna. Dependence of $\Delta$L$_{2}$ on the temperature
variation of the fork, $\Delta$T\,(steel), on the temperature variation
of the ambient air, $\Delta$T\,(amb\,air), and the variation of the wind 
speed, $\Delta$\,v (from \cite{Mangum2006}).}
\label{fig:vertex-corr-t-l2}
\end{figure}

\begin{figure}
\resizebox{\hsize}{!}{
\includegraphics[angle=-90]{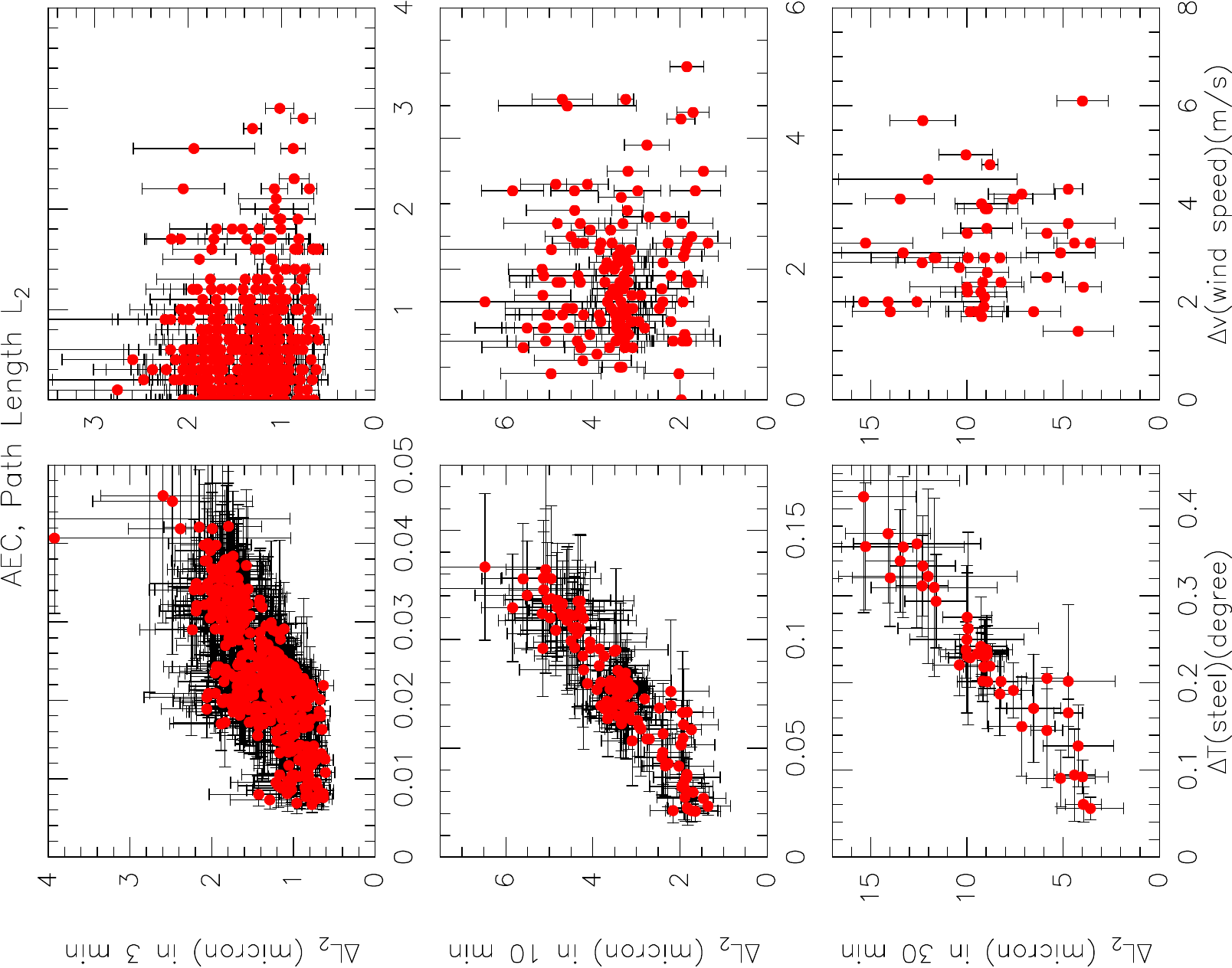}}
\caption{AEC antenna. Dependence of $\Delta$L$_{2}$ on the temperature
variation of the fork, $\Delta$T\,(steel) and the variation of the wind speed,
$\Delta$\,v. Measurements of the ambient air temperature were not
available (from \cite{Mangum2006}).}
\label{fig:aec-corr-t-l2}
\end{figure}

In order to understand the origin of the path length variations, and
for purposes of predictions, we have searched for correlation of the path 
length changes with the steel temperature of the pedestal and the fork, the 
ambient air temperature, and the wind speed. The correlation of the fork arm
path length change $\Delta$L$_{2}$ is shown for the VertexRSI antenna in 
Fig.~\ref{fig:vertex-corr-t-l2}, and for the AEC antenna in
Fig.~\ref{fig:aec-corr-t-l2}. As to be expected, the correlation of
the path length 
change $\Delta$\,L$_{2}$ with the change of the fork steel temperature is 
good, and usable for prediction.  On the VertexRSI antenna, a similarly useful 
correlation was found for the path length of the pedestal (L$_{1}$).
Also as expected, there is no significant correlation between path
length variation and ambient air temperature or wind speed variation.

\begin{figure}
\resizebox{\hsize}{!}{
\includegraphics{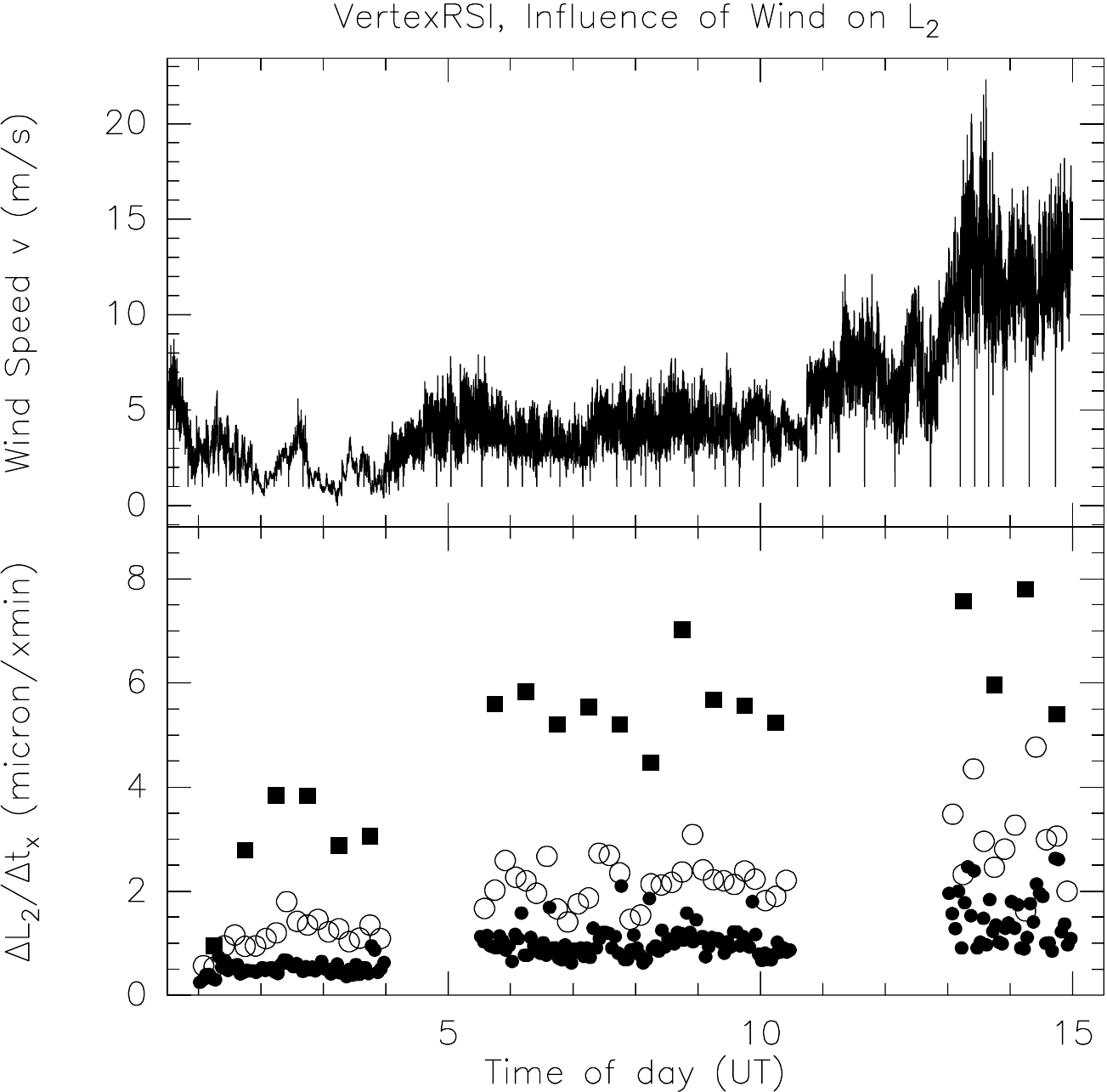}}
\caption{VertexRSI antenna. Correlation of path length variation 
$\Delta$L$_{2}$ with wind speed for 3 (dots), 10 (open circles), and
  30 (squares) minute intervals.}
\label{fig:corr-w-l2}
\end{figure}

A dedicated investigation of path length changes with wind speed was difficult
because of timing. For one particular day, with changes of the wind 
speed (v) from $\sim$\,5\,m/s to $\sim$\,15\,m/s, the path length variation 
$\Delta$L$_{2}$ (fork) of the VertexRSI antenna is shown in
Fig.~\ref{fig:corr-w-l2}. The data seem to indicate an increase of
$\Delta$L$_{2}$  
with an increase of the wind speed. However, the measured variation of L$_{2}$
is tolerable even for the specified highest wind speed of observation with 
the interferometer array, of v = 9\,m/s. There are no data for the AEC antenna.

\begin{figure}
\resizebox{\hsize}{!}{
\includegraphics{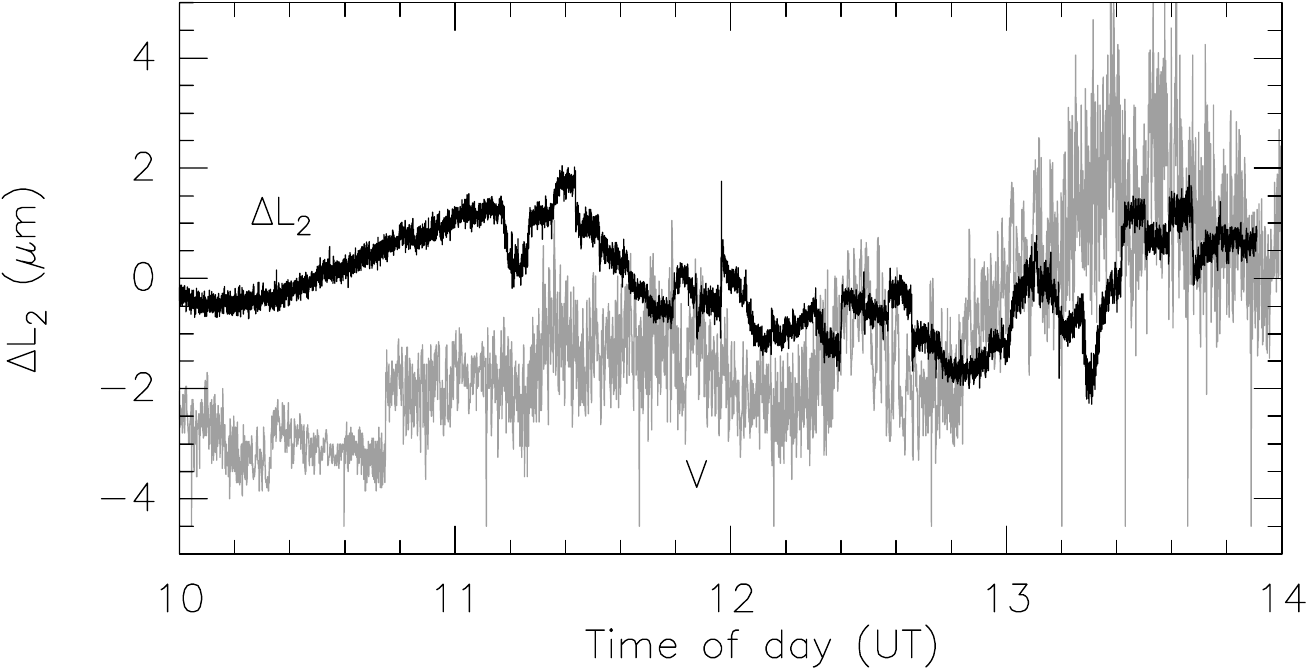}}
\caption{VertexRSI antenna. Correlation of path length variation 
$\Delta$L$_{2}$ (bold line) with wind speed for v$_{wind}$ = 0--20
m/s for a typical four hour observation period.} 
\label{fig:windL2}
\end{figure}

During the measurements shown in Figure~\ref{fig:corr-w-l2} a gradual
change of the fork temperature occured as well which produced a drift
of L$_2$ with time. This effect is not considered in the construction of
Figure~\ref{fig:corr-w-l2}.  A detailed view of the wind speed and the
path length variation $\Delta$L$_2$ with a linear temperature induced
drift of L$_2$ removed is shown in Figure~\ref{fig:windL2}. It is seen
that the path length $\Delta$L$_2$ hardly changes by $\pm 2$ $\mu$m
although the gusty wind changes velocity between 5 and 20 m/s. The
path length L$_2$ is insensitive to the gusts of the wind. The
apparent increase in path length variation $\Delta$L$_2$ with wind
speed, shown in Figure~\ref{fig:corr-w-l2}, is probably due to to a
faster temperature change of the fork at higher wind speed (convective
cooling) than a mechanical effect of the wind.

\subsection{Influence of Material Thermal Properties on $L_3$}
\label{pathtempl3}

\begin{figure}
\resizebox{\hsize}{!}{
\includegraphics{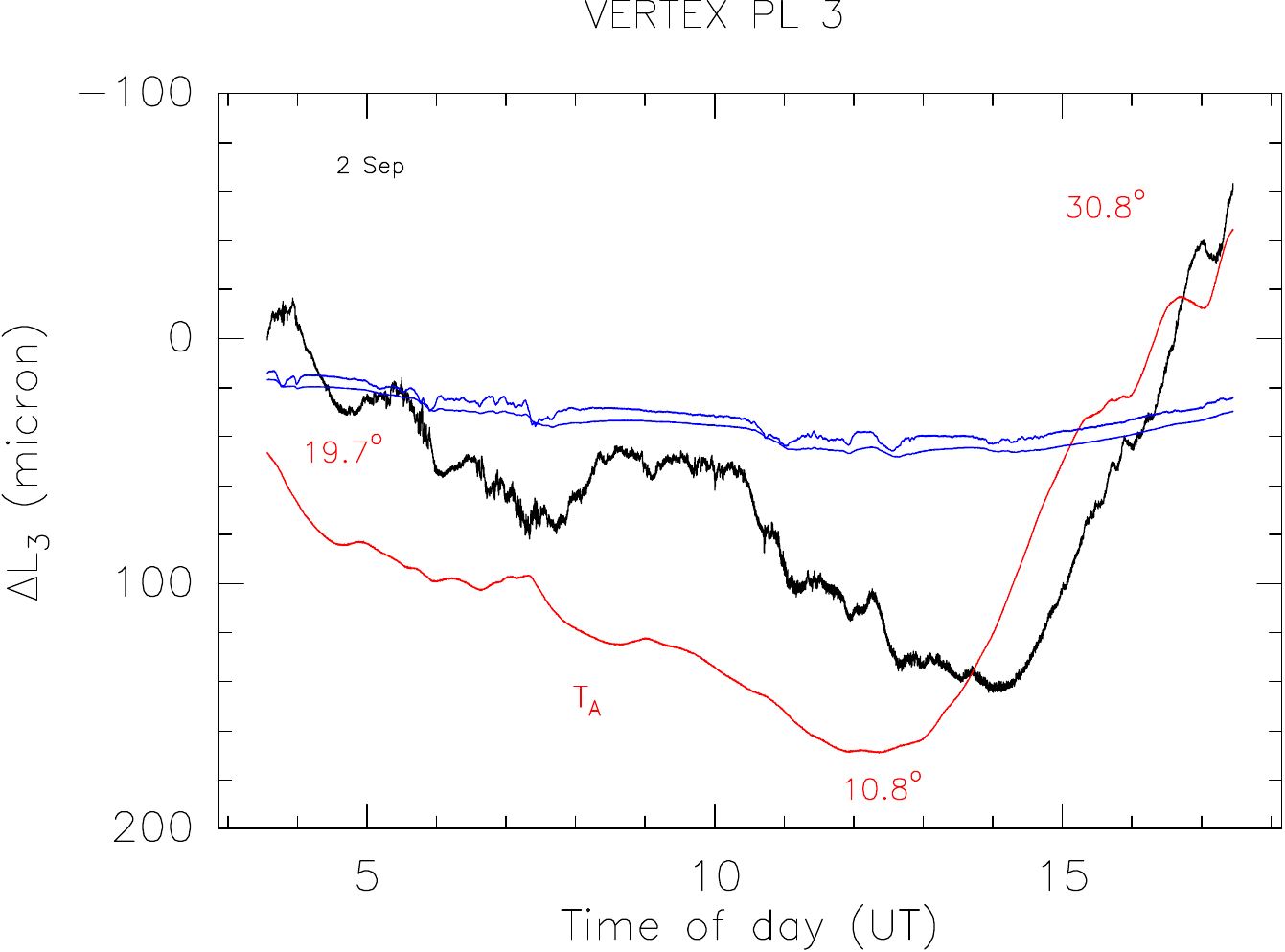}}
\caption{VertexRSI antenna. Measured path length variation $\Delta$L$_{3}$:  
black line; measured ambient air temperature: T$_A$; the air temperature 
at the beginning and the end of the measurement is given in the figure; 
temperature of the quadripod near the subreflector: parallel lines
(same scale of the temperatures).} 
\label{fig:path-l3}
\end{figure}

On both antennas, the prototype subreflector was connected to the quadripod by
a 50\,cm long aluminum tube. The large thermal dilatation of this tube 
($\sim$\,25\,$\mu$m/m/K), due to some extent to the asymmetric
influence of the ambient air (wind) and due to asymmetric solar
radiation, disturbed the path length measurements of L$_{3}$.  However, 
as shown in Fig.~\ref{fig:path-l3}, the measurements were corrected for 
this effect, as well as possible. The large scatter of $\Delta$L$_3$
in Figs.~\ref{fig:vertex-pl-summary} and \ref{fig:aec-pl-summary} is
due, to some extent, to an asymmetric influence of the ambient air
temperature (wind) and due to asymmetric solar illumination.
However, on the final antennas the use of 
aluminum should be avoided for essential structural parts which affect the 
path length. The measurement on the AEC antenna is similar.

\subsection{Path Length Predictions from Steel Temperature Measurements}
\label{pathtemppredict}

\begin{figure}
\resizebox{\hsize}{!}{
\includegraphics{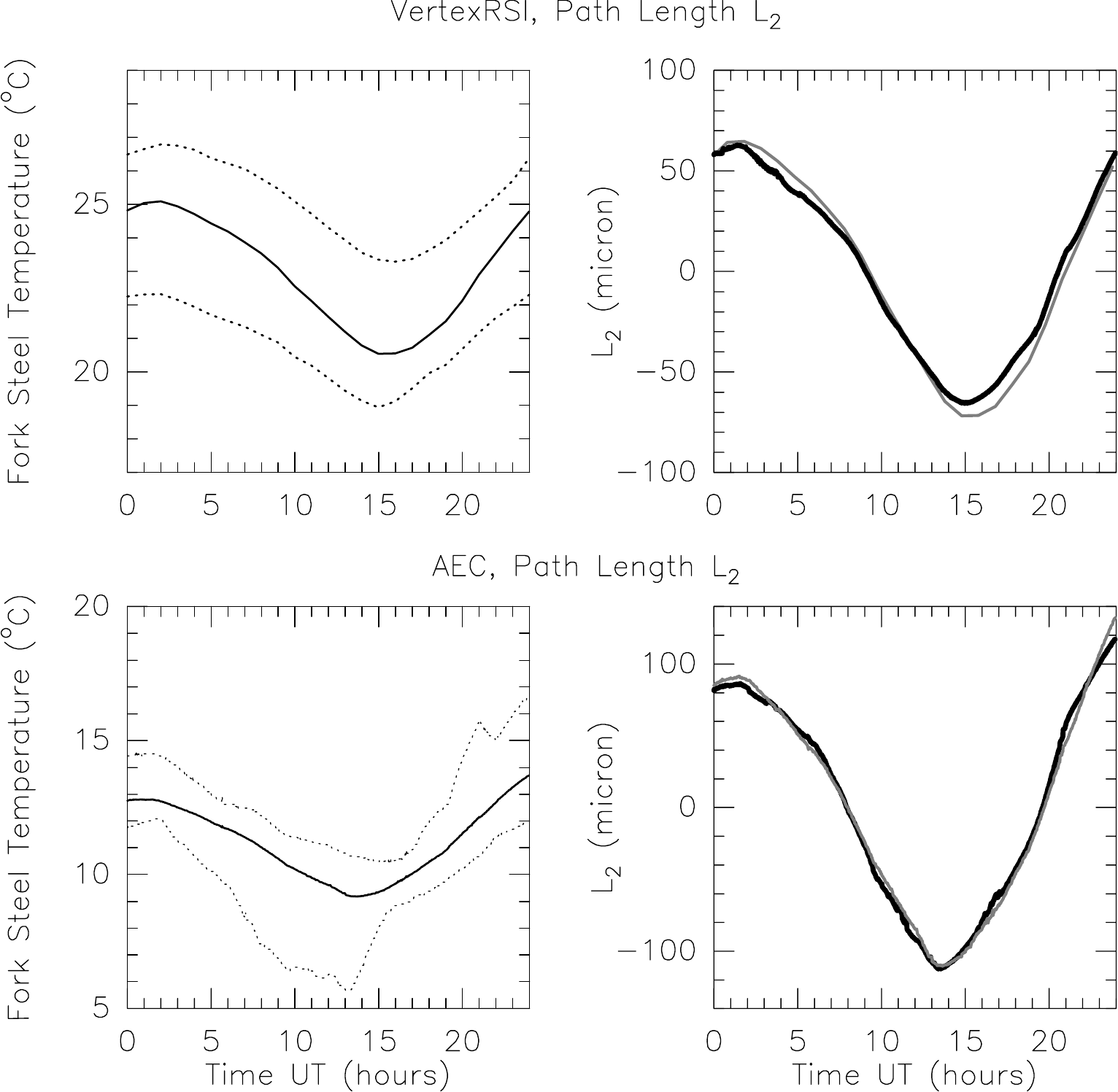}}
\caption{VertexRSI antenna and AEC antenna. Agreement between the path 
length variation $\Delta$L$_{2}$ measured with the API laser interferometer
and FEM--calculated from the measured temperatures (14 sensors). (a) average 
temperature of the fork arm steel derived from 14 sensors: solid line, maximum
and minimum fork arm steel temperature: dashed lines. (b) measured path length
variation: black line, path length variation calculated from the FEM using the
average temperature: grey line. Similar results for other days (from
\cite{Mangum2006}).} 
\label{fig:path-fem}
\end{figure}

Both antennas have temperature sensors installed on the steel walls inside the
fork arms. From these recordings we have derived the average temperature of 
each fork arm, and the maximum and minimum temperature, as shown in
Fig.~\ref{fig:path-fem}. The measured
temperature distribution was used in the finite element models (FEM)
of the fork arms to 
calculate the coresponding thermal dilatation $\Delta$\,L$_{2}$. Comparing the
calculated thermal dilatation with the path length variation measured directly
with the API, we find very good agreement as shown in
Fig.~\ref{fig:path-fem}. From this we conclude that the path length
variations of 
the steel parts can be predicted, to a high degree of accuracy, from 
representative temperature measurements used in the finite element model, or 
an empirical relation. The final number of temperature sensors of the pedestal
and the fork arms can, however, be reduced to 3 or 4, per component.  

The daily temperature variation of the fork arms and the corresponding path 
length variation $\Delta$L$_{2}$, both shown in Fig.~\ref{fig:path-fem}, 
can be approximated with good accuracy by sine--functions. This holds also 
for the ambient air temperature, at least as measured at the VLA site during 
most of the time of the tests. When adopting a sine--function L$_{2}$(t) = 
L$_{\rm o}$ + $\Delta$L$_2\sin(\omega\,t)$, with $\omega$ = 2\,$\pi$/24\,h and
t = time, the path length variation $\Delta$\,L$_{2}$ at a 3,\,10,\,and 30 
minute time interval can be derived by differentiation of the function 
L$_{2}$(t). The smallest variation of L$_2$(t) occurs around the
temperature minimum and maximum of the ambient air (T$_A$) and of the
fork steel (T$_F$).  From weather data at the site and temperature
measurements we find that T$_F\simeq$ T$_A +$3h.  As evident from
Fig.~\ref{fig:path-fem}, the maxima and minima of the fork arm
temperature (and the ambient air temperature) occur around 0h and 15h
UT (7h and 17h local time) but not late at night or at noon.

The result of the differentiation of L$_2$(t) is shown by solid lines in
Fig.~\ref{fig:vertex-pl-summary} and \ref{fig:aec-pl-summary}, which
gives an explanation of the double peaked form of the measured variations. 
The minima of $\Delta$\,L$_{2}$ occur around sunrise (around 7h local)
and sunset (around 17h local) where the temperatures go through
maximum and minimum and show the smallest change with time.

As evident from Fig.~\ref{fig:path-fem}, the total daily path length
variation of the fork 
arms is of the order of 100\,$\mu$m to 200\,$\mu$m, as fully understandable, 
and unavoidable, from the height of the fork arms, the thermal properties of
steel, the actual temperature variation of the steel, and the solar 
illumination. The relevant information for operation of the interferometer
array is however contained in the values for the 3, 10, and 30 minute
time intervals.

\subsection{Path Length Variations During Antenna Motion}
\label{pathtmotion}

The ALMA interferometer will use sidereal tracking, OTF mapping, and 
FSW motions between source and calibrator. The OTF, and in particular the FSW 
motions, involve high accelerations of the antenna which may affect the path 
length stability. The path length variations $\Delta$L$_{1}$, $\Delta$L$_{2}$ 
and $\Delta$L$_{3}$ were measured under the following motions of the antennas:
(1) sidereal tracking, as a combined motion in AZ and EL direction; (2) OTF 
motion of 1$^{\rm o}$\,AZ by 1$^{\rm o}$\,EL, at 0.05$^{\rm o}$/s; and (3) 
FSW motion of 1$^{\rm o}$\,AZ by 1$^{\rm o}$\,EL, at 6$^{\rm o}$/s in AZ and 
3$^{\rm o}$/s in EL.

\begin{figure}
\resizebox{\hsize}{!}{
\includegraphics{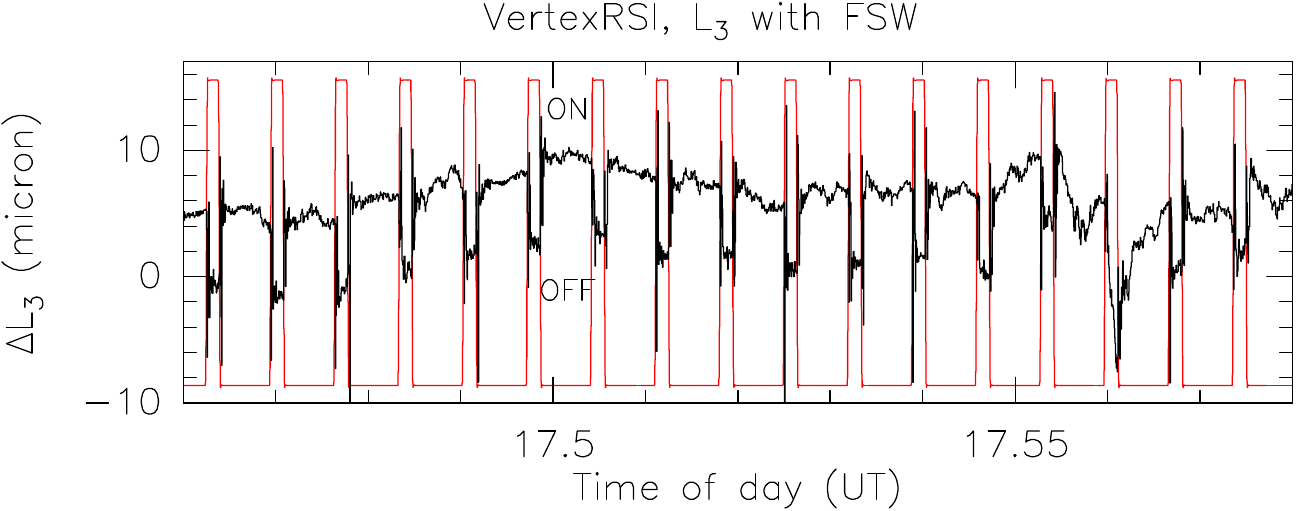}}
\caption{VertexRSI antenna. Path length variation $\Delta$L$_{3}$ (Invar cone 
to subreflector) during Fast--Switching motion.  The switching cylcle is shown
by the step line, the path length changes by the black line. The ON (10s) and 
OFF (2s) position is indicated (from \cite{Mangum2006}).}
\label{fig:vertex-path-motion-l3}
\end{figure}

\begin{figure}
\resizebox{\hsize}{!}{
\includegraphics{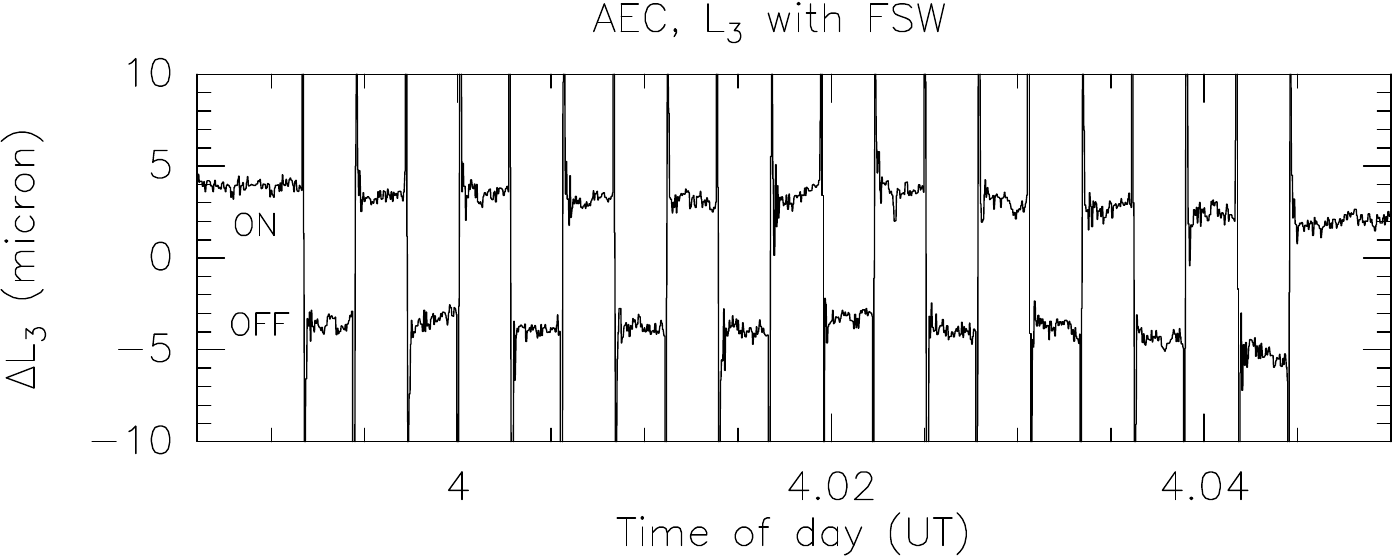}}
\caption{AEC antenna. Path length variation $\Delta$L$_{3}$ (reflector vertex 
hole to subreflector) during Fast--Switching motion.  The ON (10\,s) and OFF 
(10\,s) position is indicated (from \cite{Mangum2006}).}
\label{fig:aec-path-motion-l3}
\end{figure}

For both antennas, the variations of the path lengths L$_{1}$, L$_{2}$, and
L$_{3}$ are within $\pm$\,2\,$\mu$m for sidereal tracking and OTF motion.
For FSW motion, with the highest acceleration at the subreflector position
of the quadripod, the path length measurements L$_{3}$ are shown in
Fig.~\ref{fig:vertex-path-motion-l3} and 
\ref{fig:aec-path-motion-l3}. Again, for both antennas the path length 
variation $\Delta$L$_{3}$ at the ON and OFF position is within 
$\pm$\,3\,$\mu$m.

\subsection{Influence of Gravity on $L_3$}
\label{l3grav}

\begin{figure}
\resizebox{\hsize}{!}{
\includegraphics{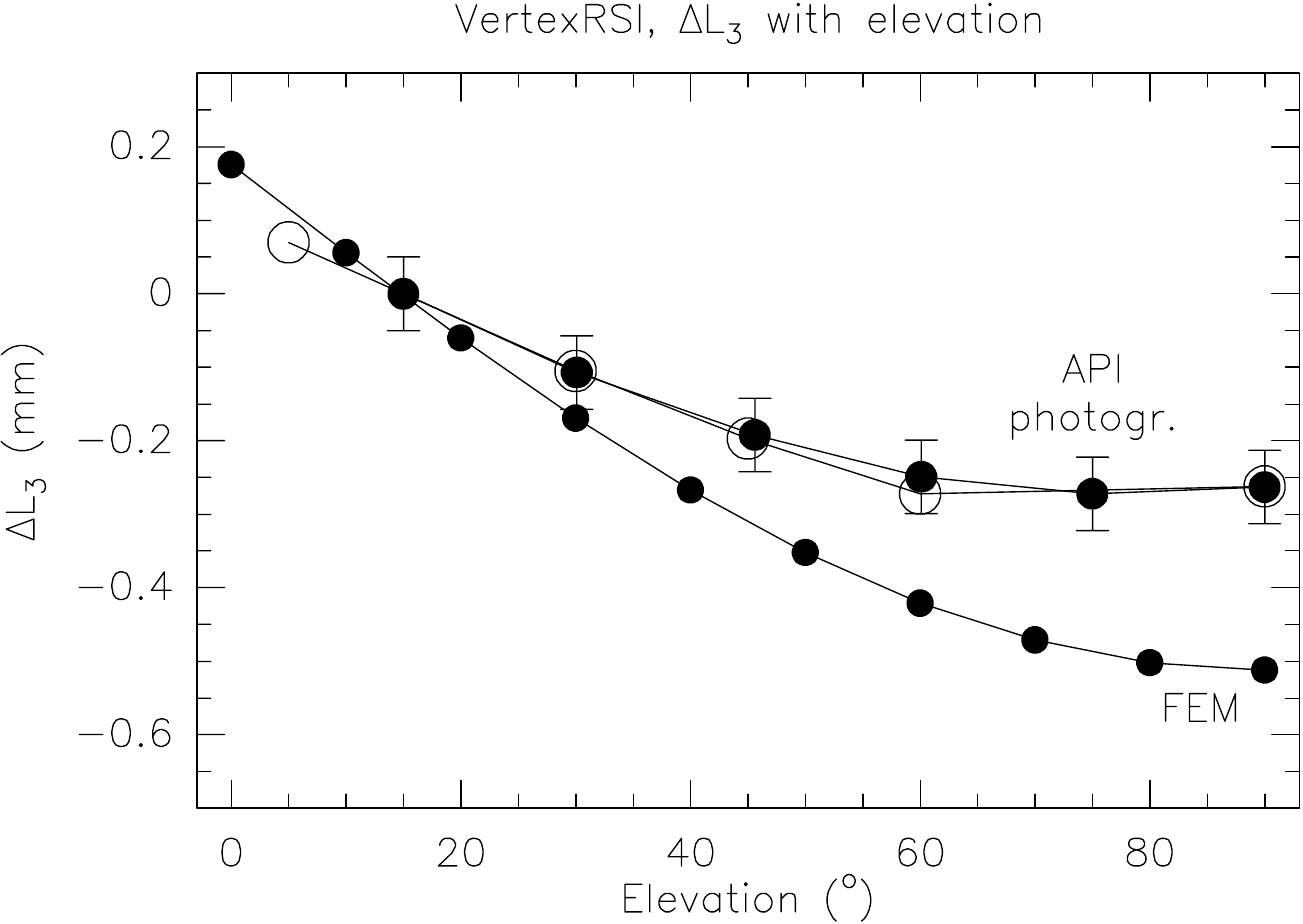}}
\caption{VertexRSI antenna. Path length variation $\Delta$L$_{3}$ as
function of elevation of the reflector, measured with the API (dots)
and by photogrammetry (open circles). The curve indicated FEM is the
calculated variation (from \cite{Mangum2006}).} 
\label{fig:vertex-el-l3}
\end{figure}

\begin{figure}
\resizebox{\hsize}{!}{
\includegraphics{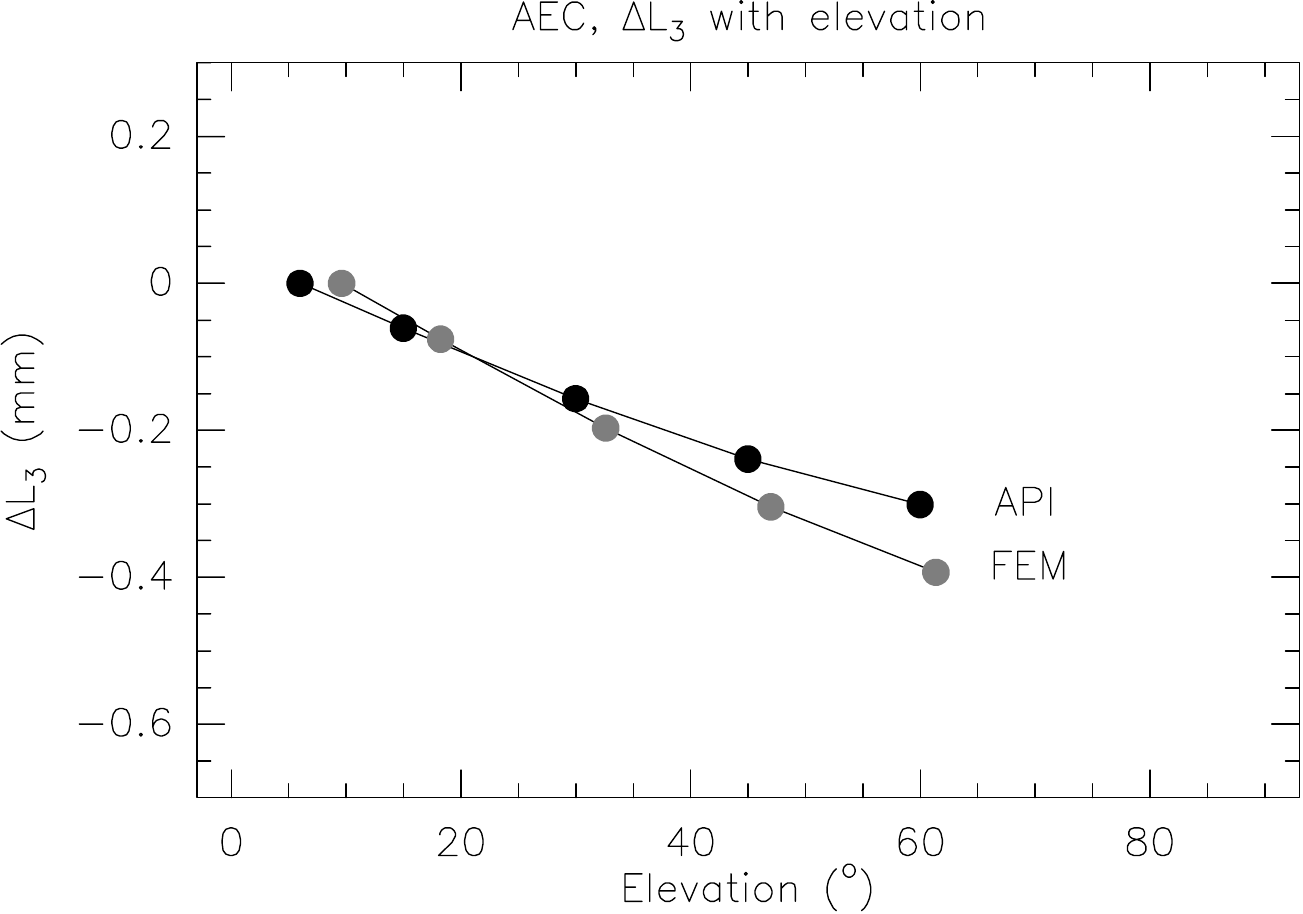}}
\caption{AEC antenna. Path length variation $\Delta$L$_{3}$ measured
  (API) as a function of elevation of the reflector. The curve
  indicated FEM is the calculated variation (from \cite{Mangum2006}).}
\label{fig:aec-el-l3}
\end{figure}

Because of gravity induced deformations, on both antennas the path length
L$_{3}$ will change with elevation of the reflector. The major part of
this change is due to the influence of gravity, and is predictable and
repeatable.  To measure this effect, 
the laser emitter of the API was installed on a mount (especially constructed
for the VertexRSI antenna; the laser tracker platform on the AEC antenna) at 
the reflector vertex, the retro--reflector was installed on the subreflector. 
The measurements of the L$_{3}$ path length variation as function of elevation 
are reliable because the measurements are insentive to a small tilt of the
laser mount relative to the antenna structure, and hence of the laser
beam reflected in the retro--reflector. The 
antennas were tipped in elevation (E) in steps of 15$^\circ$, between 
15(5)$^\circ$ and 90$^\circ$ elevation, and the variation of 
$\Delta$L$_{3}$(E) was recorded, as shown in Fig.~\ref{fig:vertex-el-l3} 
and \ref{fig:aec-el-l3}. This path length variation as predicted from a FEM 
calculation is inserted in the figures. Note in Fig.~\ref{fig:vertex-el-l3} 
that the change in subreflector position as measured by photogrammetry (GSI, 
November 2002) agrees well with the API measurement. 

\subsection{Gravitational Deformation of the AEC BUS}
\label{aeggrav}

\begin{figure}
\resizebox{\hsize}{!}{
\includegraphics{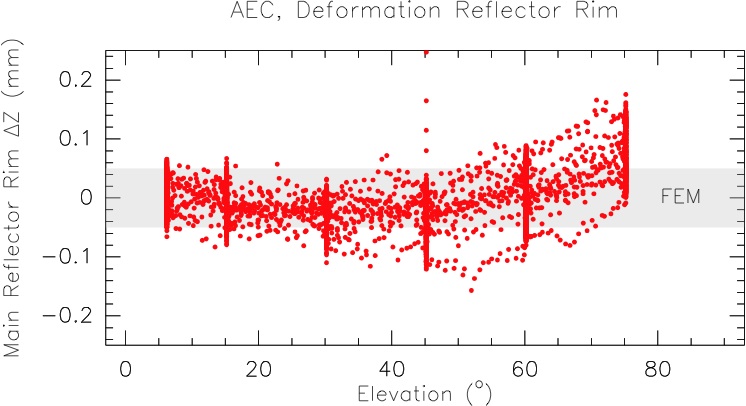}}
\caption{AEC antenna. Measured deformations of the reflector rim,
  lower half of the reflector.  The antenna was halted at the
  elevations 5, 15, 30, 45, 60, and 75 degrees. The FEM calculation
  predicts values within the indicated gray band.}
\label{fig:aec-deflect-rim}
\end{figure}

Any gravity--induced deformation of the main reflector has an influence on 
the path length (L$_4$), although it can be corrected for since the
deformation can be calculated from the FEM. Because of the stable laser 
tracker mount installed on the AEC antenna, a deflection measurement of the 
reflector rim was made with the QD. The laser emitter was installed on the 
mount at the reflector vertex, the detector was installed consecutively at 
several positions on the reflector rim (lower section). The deviation of the 
laser beam on the detector was measured while the reflector was tipped in
elevation. The result of the measurement is shown in
Fig.~\ref{fig:aec-deflect-rim} together with the prediction of the FEM
calculation.

\subsection{The API5D System as a Metrology Device}
\label{apimet}

\begin{figure}
\resizebox{\hsize}{!}{
\includegraphics{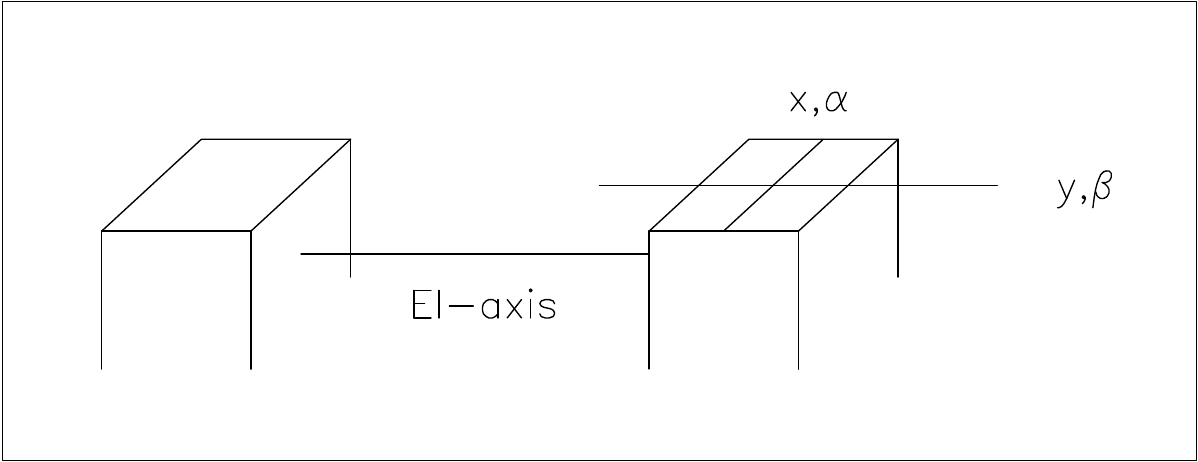}}
\caption{Orientation of the displacement directions (x,y) and the tilt
  directions ($\alpha,\beta$) for the full API5D measurements of the
  fork arm structure.}
\label{fig:apiorient}
\end{figure}

\begin{figure}
\resizebox{\hsize}{!}{
\includegraphics{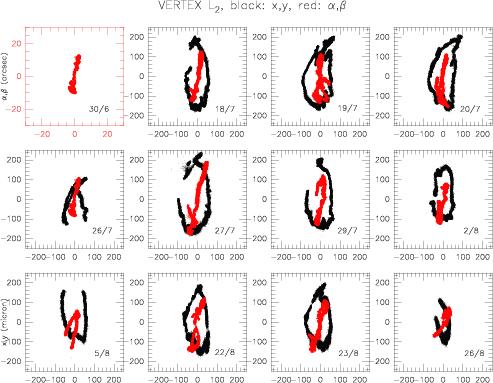}}
\caption{Measured displacements ($\Delta x,\Delta y$) (black) and
  tilts ($\Delta\alpha,\Delta\beta$) (gray) for several days as
  indicated.}
\label{fig:vertex-l2xy}
\end{figure}

The full five-dimensional capacity of the API5D system has been used
to measure the displacements ($\Delta x,\Delta y$) and the tilts
($\Delta\alpha,\Delta\beta$) of a location close to the elevation
bearing(s), and to use this information for the elevation axis
pointing error (nodding error).  On the VertexRSI antenna a few API5D
recordings were made (while using the $\Delta z$ information) with the
laser emitter placed on the transverse of the fork and the detector
installed just below the elevation bearing (see
Fig.~\ref{fig:pathlength-layout}, L$_2$).  The orientation of the
(x,y,$\alpha$,$\beta$) axes and rotations are shown in
Fig.~\ref{fig:apiorient}, the recordings of ($\Delta x, \Delta y,
\Delta\alpha, \Delta\beta$) during several days are shown in
Fig.~\ref{fig:vertex-l2xy}.  A rather regular daily excursion is seen,
of the order of $(\Delta x, \Delta y, \Delta\alpha, \Delta\beta)
\simeq (\pm 100 \mu m, \pm 200 \mu m, \pm 5 arcsec, \pm 15 arcsec)$.
For a length of the fork arm of L$_2 \simeq 3$m, it is found that
$\Delta x^\prime \simeq L_2 \Delta\alpha \simeq 70 \mu m$ and $\Delta
y^\prime \simeq L_2 \Delta\beta \simeq 200 \mu m$, apparently in
agreement with the measured values ($\Delta x,\Delta y$).  This
agreement also indicates that the ($\Delta\alpha, \Delta\beta$) data
are probably not independent.

While it is plausible that the measured effect is due to temperature
variations, the FEM calculations using the measured temperatures of
the fork arm (see Fig.~\ref{fig:path-fem}) yielded dilatations and
rotations of the steel plates which did not match the measurement.
The measured behaviour is also not seen in the pointing model.  Our
result may be in the same line of similar difficulties encountered in
temperature and inclinometer measurements made on larger alidade
structures (\cite{Ambrosini1996}).

Since the data of Figure~\ref{fig:vertex-l2xy} show a clear
repeatability, over several days, it still is worthwhile to pursue an
interpretation since it may open the possibilty to measure pointing
changes along the elevation axis and  perpendicular to the elevation
axis (nodding error). The higher stability of the API system may be of
advantage with respect to inclinometers often having drifts.

\section{Thermal Behaviour}

\subsection{Distribution of Temperature Sensors}
\label{tsensordist}

\begin{figure}
\resizebox{\hsize}{!}{
\includegraphics{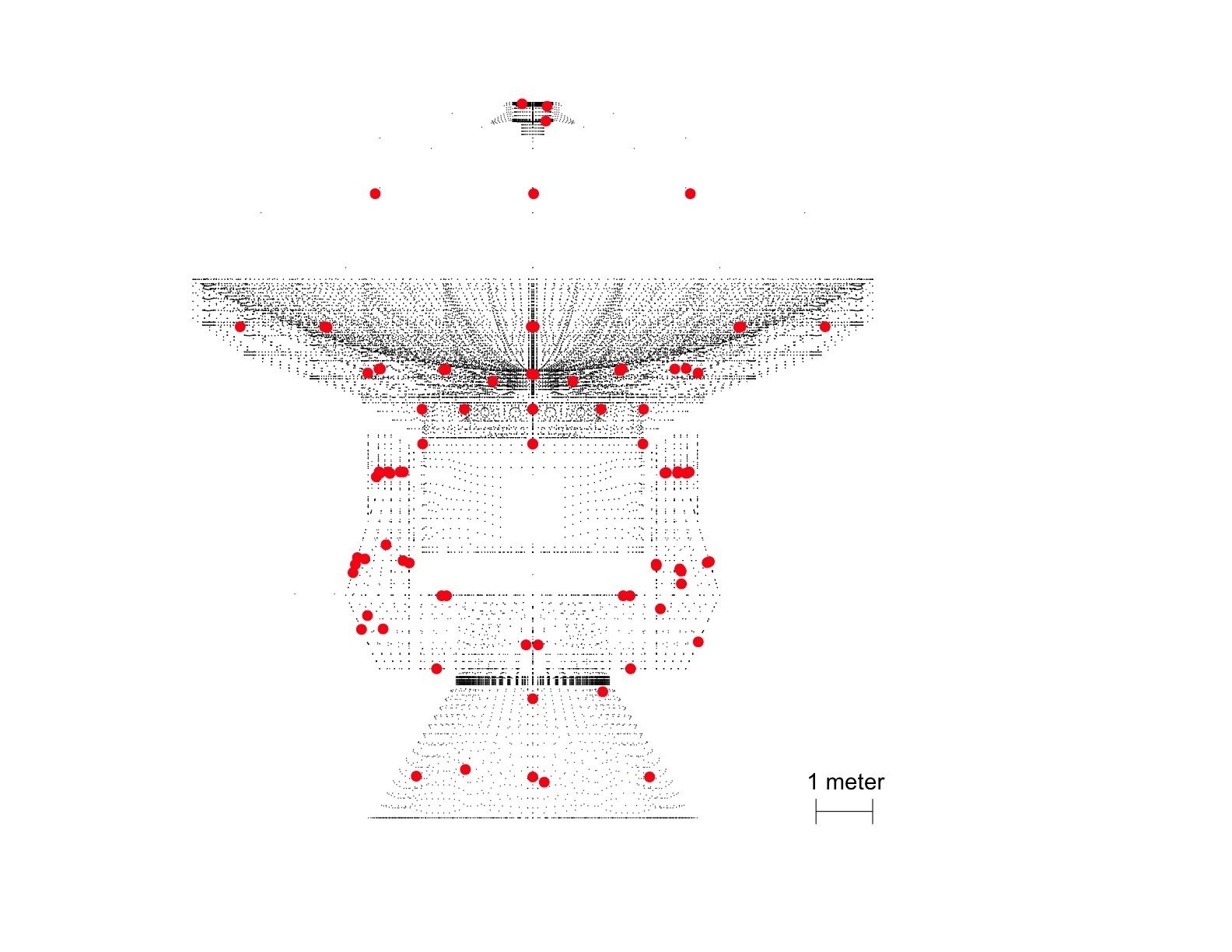} \\
\includegraphics{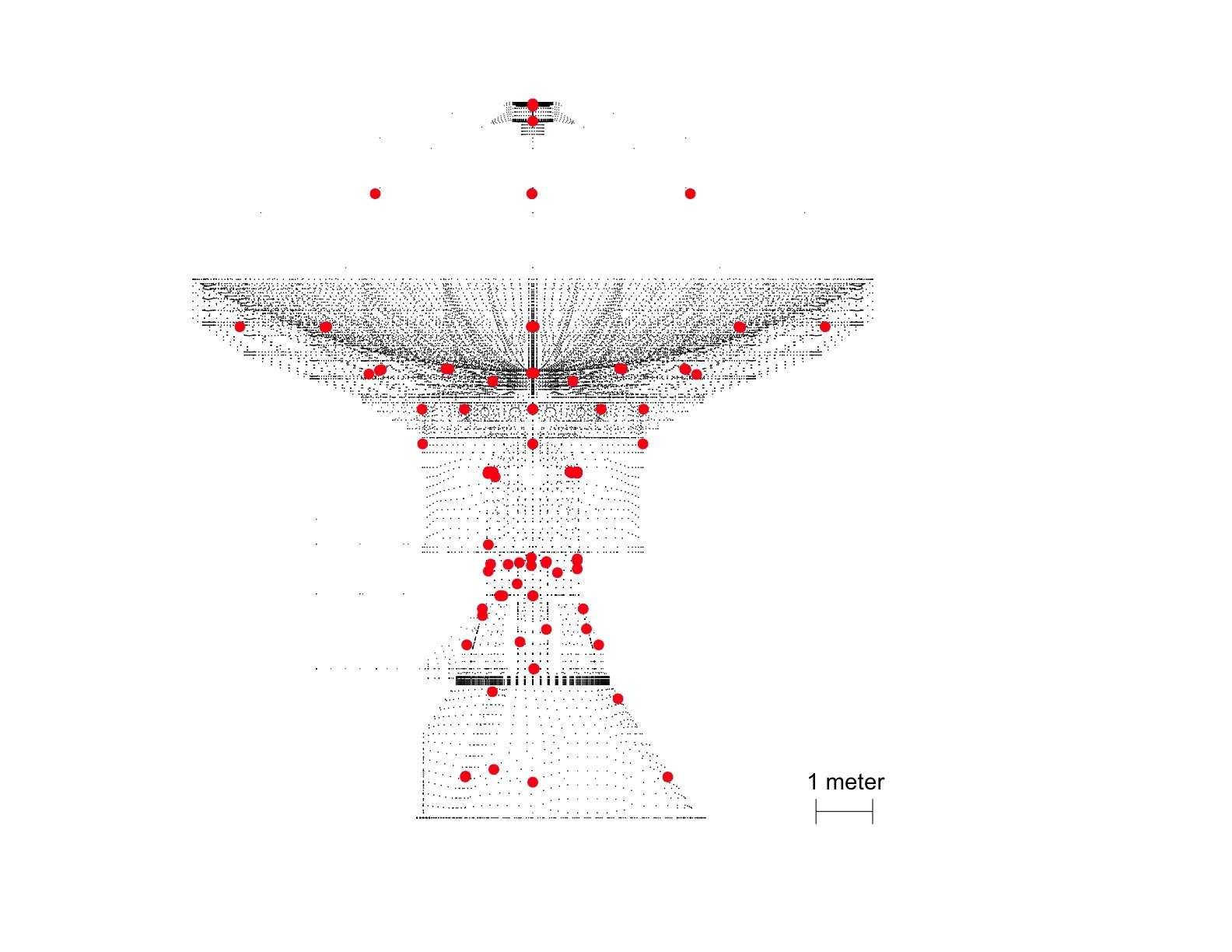}}
\caption{VertexRSI antenna. Distribtion of temperature sensors (big
  dots). The small dots are nodes of the FEM.} 
\label{fig:vertex-tsensor}
\end{figure}

\begin{figure}
\resizebox{\hsize}{!}{
\includegraphics{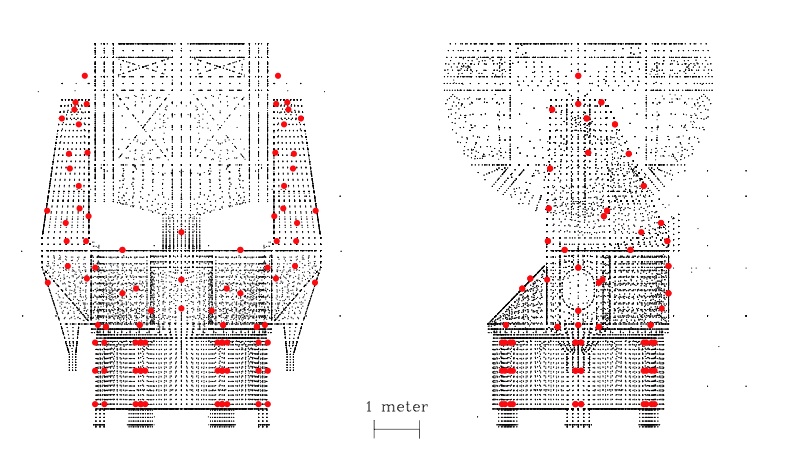}}
\caption{AEC antenna. Distribution of tempearture sensors (big dots).
  The small dots are nodes of the FEM.} 
\label{fig:aec-tsensor}
\end{figure}

An investigation of the thermal homogeneity of the antennas gives insight 
into the expected deformation of these structures. The VertexRSI antenna is 
equipped with 89 PT--100 temperature sensors
(Fig.~\ref{fig:vertex-tsensor}).  The steel components of the antenna
(pedestal, fork) are insulated,
while the steel focus cabin is in addition temperature controlled. The BUS 
is supported on an Invar cone; the BUS and the quadripod are made of 
honeycomb--CFRP plates and CFRP. The pedestal and the fork of the AEC antenna 
are equipped with 101 sensors (Fig.~\ref{fig:aec-tsensor}). The
pedestal and fork is 
insulated. The focus cabin, BUS, and quadripod are made of CFRP, and were not 
fitted with thermal sensors due to the low coefficient of thermal expansion
of these components.

On the VertexRSI antenna the temperature sensors were regularly placed
to represent approximately equal volume elements of the structure.  On
the AEC antenna, the temperature sensors were part of the metrology
system, though not tested in this function.  Using the FEM in the way
described by \cite{Bremer2002}, AEC antenna temperature sensors were
placed to give optimal information about the temperature-induced
pointing errors from the fork arm supports.

\subsection{Measured Temperatures}
\label{meastemp}

\begin{figure}
\resizebox{\hsize}{!}{
\includegraphics[angle=-90]{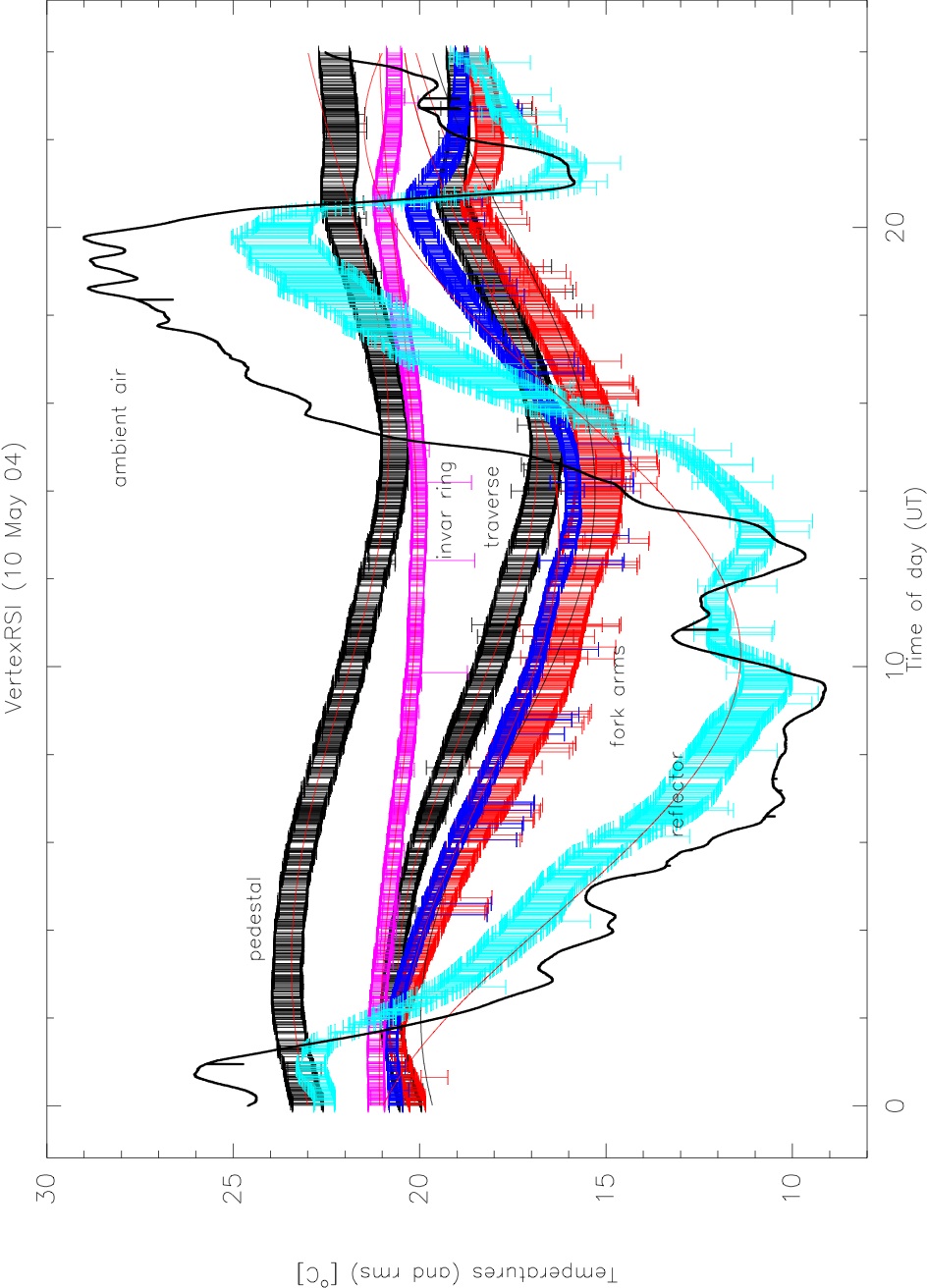}}
\caption{Measured temperatures of the VertexRSI antenna, for individual 
 components as indicated. The width of the lines is the rms--value of the 
 temperature throughout the indicated component. Sinusoidal fits are shown in 
 red. Very similar temperature distributions were measured on other days 
 during the same season.}  
\label{fig:vertex-temp-example}
\end{figure}

For example, Fig.~\ref{fig:vertex-temp-example} shows a
representative diurnal temperature distribution measured throughout
the VertexRSI antenna.  For the day shown, we unfortunately have no
measurements of the solar illumination.  It was a clear day and the
antenna parked in a stationary position looking towards the southern
horizon. We note that the 
temperature variation of the pedestal, the traverse, and the fork arms
are small. Also, the temperature variation, and the rms value, of the 
Invar cone, as found for this day and other days, is quite small. The
Invar cone is the ventilated upper ring of the focus cabin, which itself is
stabilized in temperature by active thermal control. The Invar cone supports
the BUS, and hence no print--through of thermal deformations is
expected to occur. The BUS itself shows a large daily temperature
variation, similar to that of the ambient air. However, the
rms--deviation of the temperature is small, of the order of 3$^{\rm
  o}$\,C.  A small differential thermal deformation of the BUS is
expected to occur, though being rather harmless, since the BUS
consists of 24 independent sectors, with each sector containing
several smaller compartments.

\begin{figure}
\resizebox{\hsize}{!}{
\includegraphics[angle=-90]{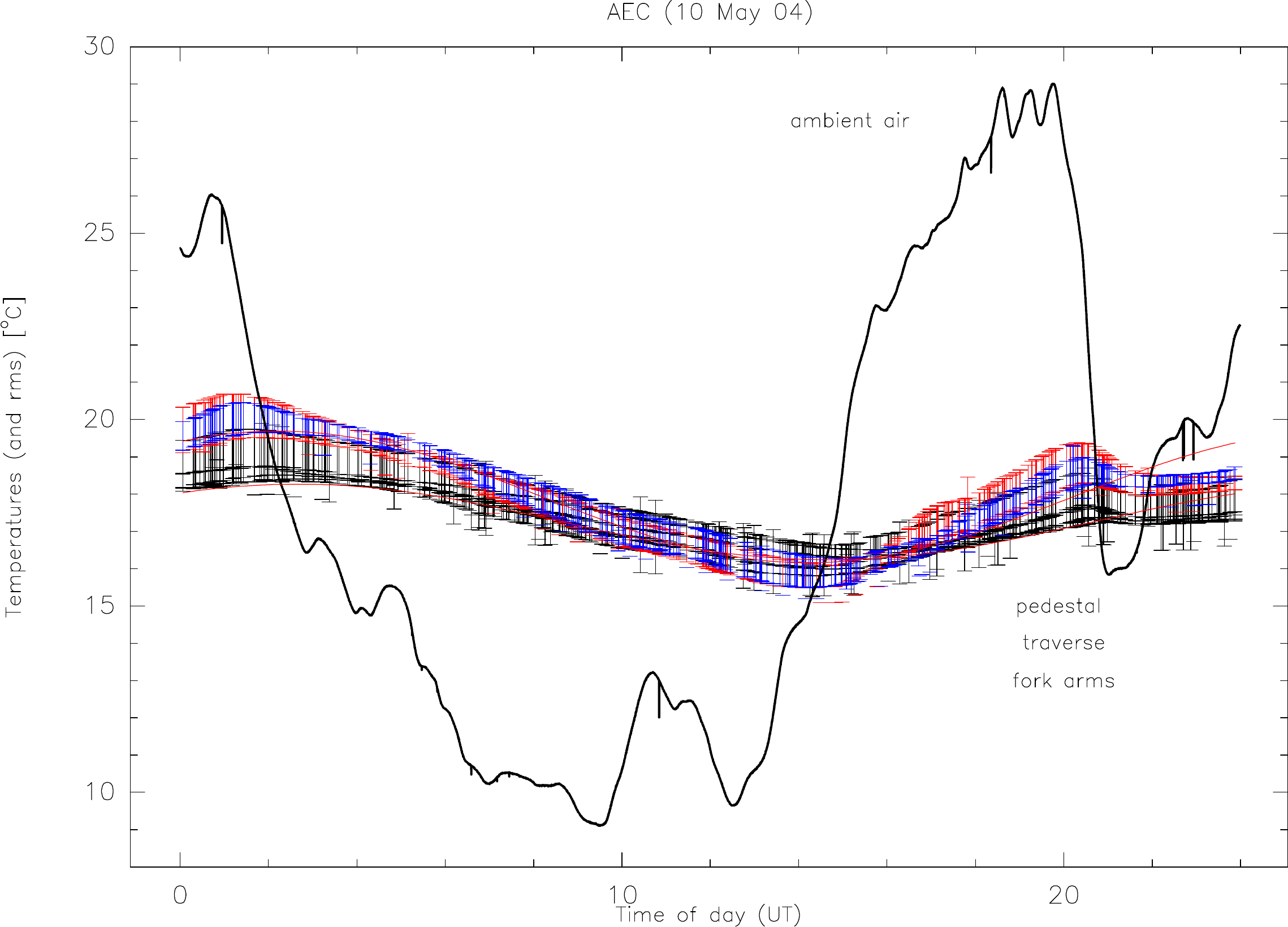}}
\caption{Measured temperatures of the AEC antenna, for individual
  components as indicated. The width of the lines is the rms--value of the 
  temperature throughout the indicated component. Sinusoidal fits are shown 
  in red. Very similar temperature distributions were measured on other days
  during the same season.}  
\label{fig:aec-temp-example}
\end{figure}

\noindent{Fig.}~\ref{fig:aec-temp-example} shows, for the same day, the
temperature distribution measured on the AEC antenna.  The temperatures of the
components are very similar, while also the temperature variations are very 
small.

\begin{figure}
\resizebox{\hsize}{!}{
\includegraphics{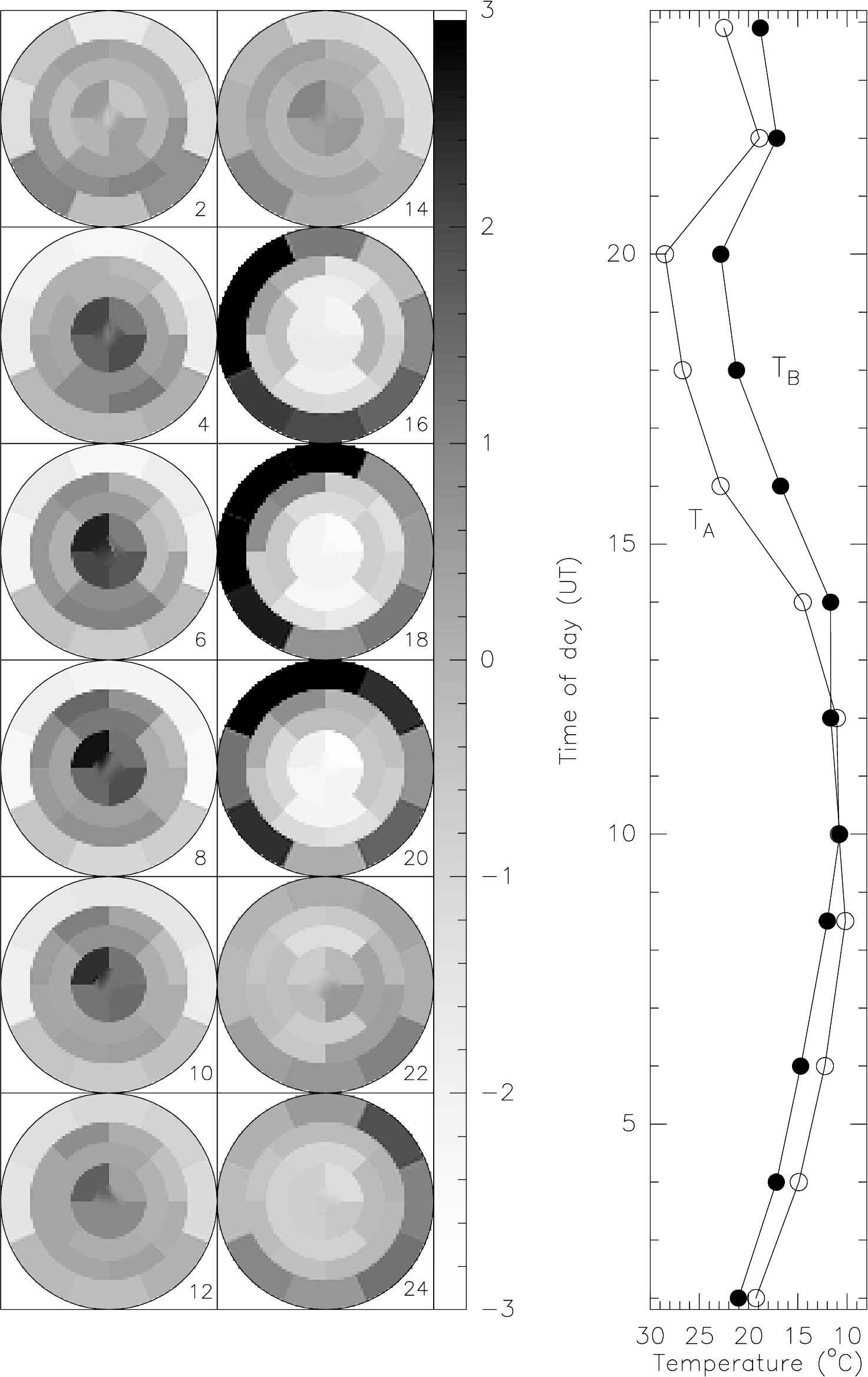}}
\caption{VertexRSI antenna. Temperature distribution of the 26 BUS elements 
monitored with sensors; stationary antenna. The left panel shows the 
deviations from the instantaneous average temperature of the BUS. The wedge 
indicates temperature in degree C. The right panel shows the average 
temperature of the BUS (T$_B$) and the temperature of the ambient air
(T$_A$).  Time in UT (2--24 hr).} 
\label{fig:vertex-temp-sun-may}
\end{figure}

The sensors of the VertexRSI antenna allow the measurement of the temperature
distribution throughout the BUS (Fig.~\ref{fig:vertex-tsensor}). From
the measurement of 26 sections of the 
BUS we determined the average temperature, i.e. T$_{\rm B}$, and deviations 
from the average $\Delta$T$_{\rm i}$ = T$_{\rm i}$ -- T$_{\rm B}$, with i = 
1,...,\,26. The values $\Delta$T$_{\rm i}$ determine the degree of thermal 
homogeneity and the residual thermal deformation of the BUS. A representative 
temperature recording is shown in Fig.~\ref{fig:vertex-temp-sun-may}.

\begin{figure}
\resizebox{\hsize}{!}{
\includegraphics{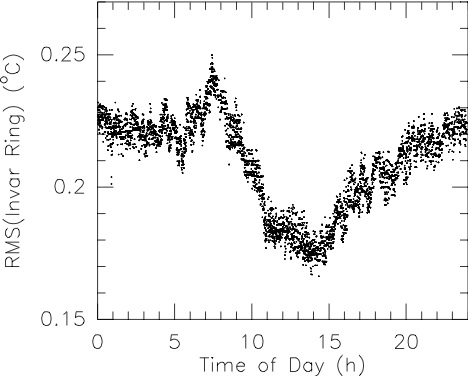}}
\caption{VertexRSI antenna. The rms value of the temperature
  homogeneity of the invar ring supporting the BUS. Data from four
  consecutive days.}
\label{fig:inv-rms}
\end{figure}

The BUS of the VertexRSI antenna is made of CFRP-plated Al-honeycomb
plates, while the BUS of the AEC antenna is made of CFRP plates. The
thermal behaviour of both BUS constructions is expected to be
comparable. However, the support of the BUS is different.  The BUS of
the VertexRSI antenna is supported on a temperature stabilized Invar
ring, while the BUS of the AEC antenna is supported on a CFRP support
structure. The temperature homogeneity of the Invar ring, expressed as
the rms value of the temperature distribution, is shown in
Figure~\ref{fig:inv-rms}, derived from measurements of four
consecutive days. The rms value is $\sim 0.25$ C all the time. The BUS
rests on a stable support. 

\subsection{VertexRSI Antenna: Influence of Direct Solar Radiation}
\label{vertexsolar}

\begin{figure}
\resizebox{\hsize}{!}{
\includegraphics{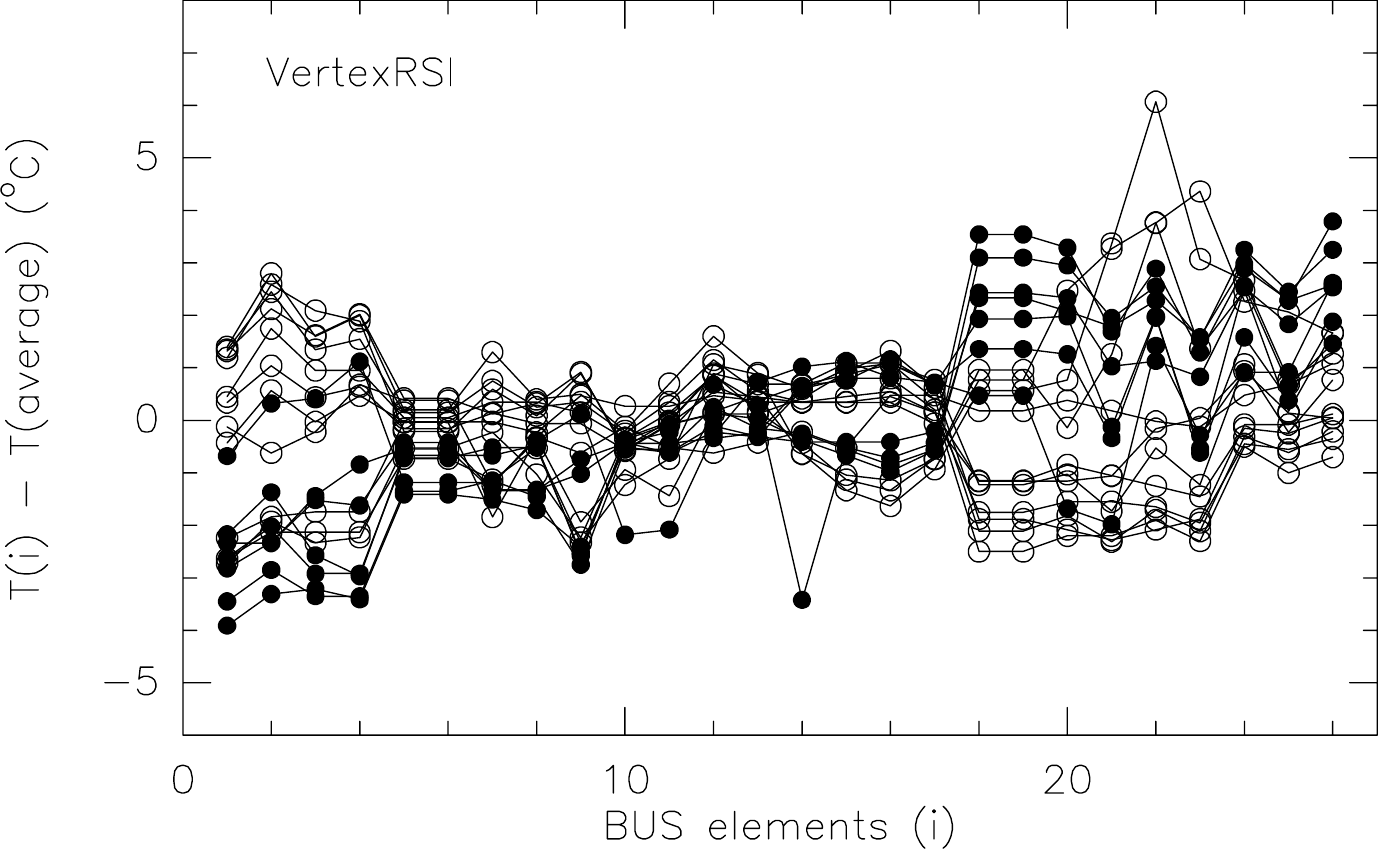}}
\caption{VertexRSI antenna. Temperature distribution of the BUS for a
  staionary antenna (open circles) facing south, and the antenna
  tracking the Sun (dots). 
  Shown are the temperature deviations of the 26 BUS elements with respect to 
  the instantaneous average temperature of the BUS. Each line, is for
  a selected hour of the day.}
\label{fig:vertex-temp-compare}
\end{figure}

The ALMA antennas may be used for observations of the Sun, or observations
very close to the Sun. A temperature measurement of the VertexRSI antenna 
tracking the Sun was made on 13 Jul 2003. The comparison of the temperature 
distribution of the 26 BUS elements for a normal day
(Fig.~\ref{fig:vertex-temp-sun-may}) and the Sun tracking condition is
shown in Fig.~\ref{fig:vertex-temp-compare}. This figure illustrates that
there is no noticable distinction in the temperature variation across
the BUS for an observation which tracks the Sun, as requested in the
antenna specifications.

\subsection{Thermal Response of Antenna Components}
\label{tresponse}

From the continuously monitored temperatures T$_{\rm i}$ of the antenna 
components [i] and the ambient air temperature T$_{\rm A}$ we determined the 
thermal response $\beta$$_{\rm i}$ using the relation $\Delta$T$_{\rm i}$ = 
$\beta$$_{\rm i}$\,$\Delta$T$_{\rm A}$, with $\Delta$T$_{\rm i}$ and 
$\Delta$T$_{\rm A}$ the daily amplitude of the respective temperature. This 
relation neglects the influence of direct solar radiation, however, the 
antennas were randomly observing the sky. For a time of $\sim$\,14 days, the 
values $\beta$$_{\rm i}$ in Table \ref{tab:insulation} indicate that
both antennas behave in a 
similar way. On the VertexRSI antenna, the invar cone shows a low thermal 
response, which together with the thermally stabilized focus cabin guarantees 
a stable base for the BUS. Using the data of
Fig.~\ref{fig:vertex-corr-t-l2} and Fig.~\ref{fig:aec-corr-t-l2}, we
are able to relate the path length variations $\Delta$\,L$_{1}$
(pedestal) and $\Delta$\,L$_{2}$ (fork) to the variation of the steel
temperature (T$_{\rm M}$) and the ambient air temperature (T$_{\rm
  A}$). The corresponding values are given in Table \ref{tab:pedfork}.    

\begin{table}[h]
\centering
\caption{Insulation Efficiency $\beta$ of the Antenna Components}
\begin{tabular}{|lccccc|}
\hline
Antenna & Pedestal & Traverse & Fork (L/R) & Invar Cone & BUS  \\
\hline
VertexRSI & 0.14 & 0.20 & 0.24/0.26 & 0.06 & 0.72 \\
AEC       & 0.13 & 0.22 & 0.22/0.14 & --   & --  \\
\hline
\end{tabular}
\label{tab:insulation}
\end{table}
\begin{table}[h]
\centering
\caption{Pedestal and Fork Arm Path Length Variation (in mm/$^\circ$\,C)}
\begin{tabular}{|lcccc|}
\hline
Antenna & Pedestal & Fork Arm & Pedestal & Fork Arm  \\
  & $\partial$L$_{1}$/$\partial$T$_{M}$ & $\partial$L$_{2}$/$\partial$T$_{M}$ &
  $\partial$L$_{1}$/$\partial$T$_{A}$ & $\partial$L$_{2}$/$\partial$T$_{A}$ \\
\hline
VertexRSI & 0.10 & 0.03  & 0.15  & 0.0075 \\
AEC       & --   & 0.025 &  --   & 0.005 \\
\hline
\end{tabular}
\label{tab:pedfork}
\end{table}

\subsection{Thermal Tilt of the Elevation Axis}
\label{eltemptilt}

The sensors of each fork arm allow the determination of the
differential thermal dilatation of the arms:

\begin{equation}
\delta L_2 = L \Delta T_l - L \Delta T_r.
\label{deltal2}
\end{equation}

This differential dilatation results in a change of the elevation axis
tilt:

\begin{equation}
\Delta\epsilon = \frac{\delta L_2}{D}
\label{deltaepsilon}
\end{equation}

\begin{figure}
\resizebox{\hsize}{!}{
\includegraphics{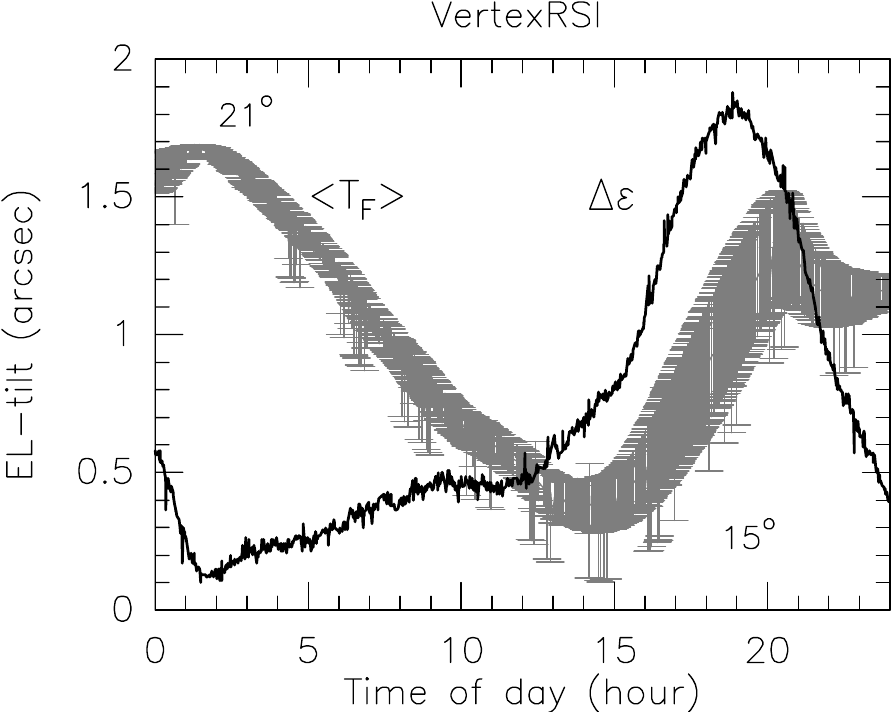}}
\caption{VertexRSI antenna.  Average temperature of the fork arms
  $<T_f>$ and temperature dispersion (width of the line).  Predicted
  axis tilt $\Delta\epsilon$.} 
\label{fig:eltilt}
\end{figure}

\noindent{in} the direction of the axis.  Although this tilt will be
recovered, and corrected, in regular pointing observations, the
magnitude $\Delta\epsilon$  gives nevertheless an indication of the
thermal stability of the fork.  For the selected day,
Fig.~\ref{fig:eltilt} shows the measured average temperature of the
fork arms $<T_f>$, with the width of the line indicating the value
$|T_f(l)-T_f(r)|$, and the tilt $\Delta\epsilon$.  There occurs a
change in $\Delta\epsilon$ with sunrise and sunset.  The change,
evaluated for approximately 14 days, is expected to be $\Delta\epsilon
\lesssim 5^{\prime\prime}$ even if the antennas of the arrays track a
target for a long time while asymmetrically illuminating the fork arm
structure.

\section{Precision of the Azimuth Bearing}

\subsection{Measurements of the VertexRSI Antenna}
\label{vertexaz}

The specification requires a \textit{2--arcsecond (rms) blind pointing
  accuracy and a 0.6--arcsecond (rms) pointing accuracy for FSW
  observations in a field of 2$^\circ$ radius}. For analysis of the
antenna pointing and the understanding of the pointing model it is
useful to know 
which part of the azimuth axis pointing error is due to the azimuth bearing. 
Accordingly, appropriate higher order terms of the azimuth angle A are 
introduced in the pointing model.

\begin{figure}
\resizebox{\hsize}{!}{
\includegraphics{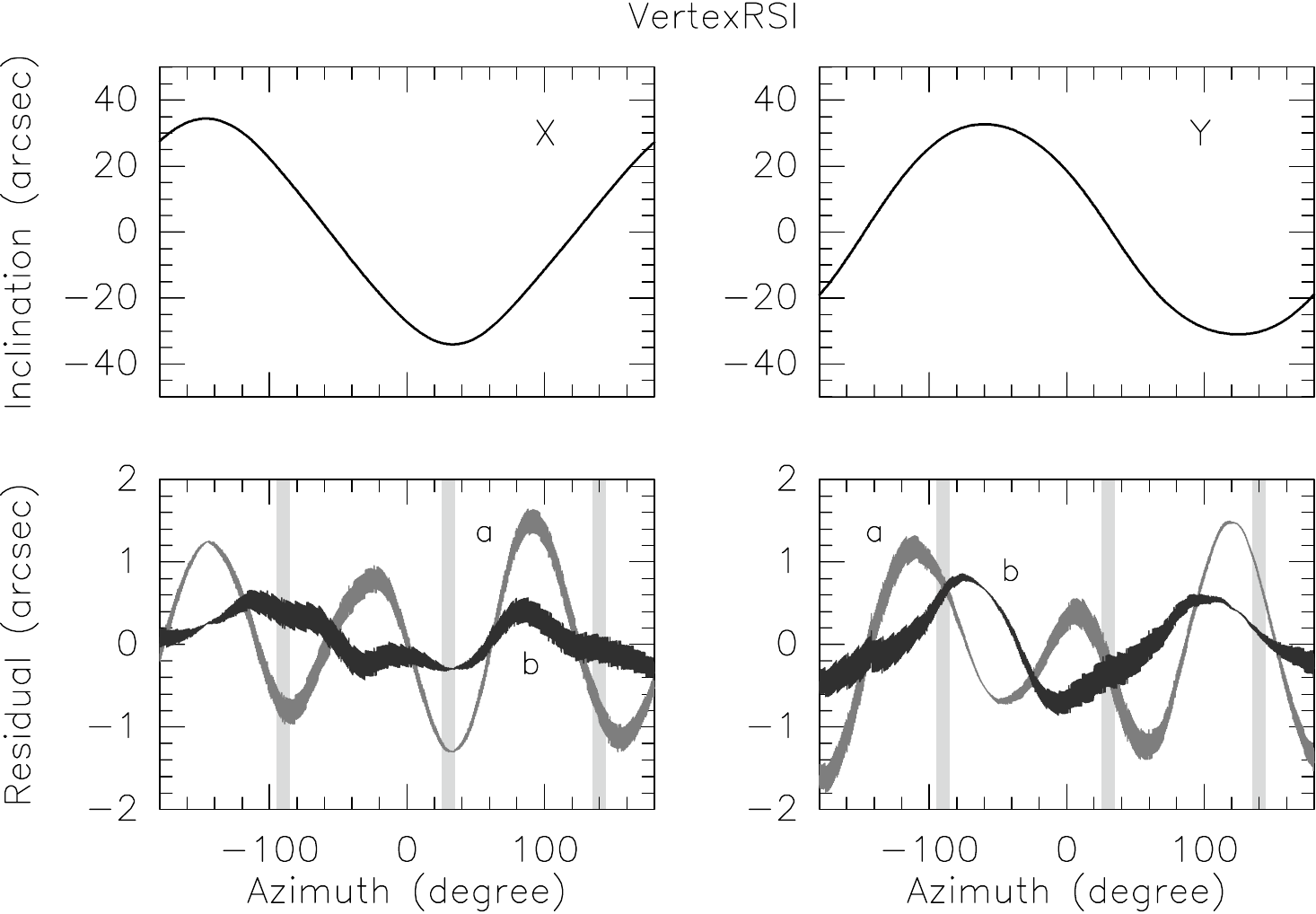}}
\caption{VertexRSI antenna, measurement of --\,180$^{\rm o}$ to 
+\,180$^{\rm o}$ AZ rotation. The left panel is the x--direction of the 
tiltmeter, the right panel is the y--direction. The curves
\textit{a} are the residuals of the best--fit a$_{0}$ +
a$_{1}$\,sin(A + b$_{1}$), the curves \textit{b} are the residuals of
the best--fit a$_{0}$ + a$_{1}$\,sin(A + 
b$_{1}$) + a$_{3}$\,sin(3\,A + b$_{3}$). The gray lines show the
location of the pedestal corners (1,2,3) (from \cite{Mangum2006}).}
\label{fig:vertex-azwobble}
\end{figure}

Measurements with the tiltmeter installed above the AZ bearing made at a slow 
speed (0.1$^\circ$/s) and $\pm$\,260$^\circ$ rotation in AZ show, for both
antennas, a single--angular sinusoidal response [sin(A)] to the inclination 
of the antenna (and foundation), and in addition a smaller three--angular 
sinussoidal response [sin(3\,A)] on the VertexRSI antenna, and an additional 
two--angular response [sin(2\,A)] on the AEC antenna. On the VertexRSI 
antenna, the three--angular wobble amounts to $\pm$\,2\,arcsec; the extrema 
of the wobble are correlated with the position of the  three--corner support 
of the pedestal, as shown in Fig.~\ref{fig:vertex-azwobble}. This
three--angular wobble is a print--through of the pedestal mount. After
elimination of the components sin(A) and sin(3\,A), the residual
repeatable deviation is 0.5\,arcsec rms.  The best--fit values of the
measurements to a$_{0}$ + a$_{1}$\,sin(A + b$_{1}$) +
a$_{3}$\,sin(3\,A + b$_{3}$) are given in Table
\ref{tab:vertex-azwobble}.  The residual curves \textit{b} in
Fig.~\ref{fig:vertex-azwobble} can be used in a lookup table for
correction.

\vskip 0.1 cm

\begin{table}[h!]
\centering
\caption{VertexRSI Antenna: Parameters of the AZ axis Inclination and 
Wobble (in brackets values remeasured after one year)}
\begin{tabular}{|cr|cr|}
\hline 
 Linear             &  response    &  Linear  & + 3\,A response \\
\hline
 a$_{0}$ &  64.49\,(53.08)$''$        & a$_{0}$ &  64.47\,(53.08)$''$    \\
 a$_{1}$ & --\,68.41\,(--\,65.02)$''$ & a$_{1}$ & --\,68.31\,(--\,64.93)$''$ \\
 b$_{1}$ &  56.66\,(59.77)$^{\rm o}$  & b$_{1}$ &  56.811\,(59.91)$^{\rm o}$ \\
         &                            & a$_{3}$ & --\,2.29\,(--\,2.35)$''$  \\
         &                            & b$_{3}$ & --\,10.84\,(--\,4.64)$^{\rm o}$  \\
\hline
 rms(residual) & 1.69\,(1.70)$''$     & rms(residual) & 0.60\,(0.50)$''$ \\
\hline
\end{tabular}
\label{tab:vertex-azwobble}
\end{table}

\subsection{Measurements of the AEC Antenna}
\label{aecaz}

\begin{figure}
\resizebox{\hsize}{!}{
\includegraphics{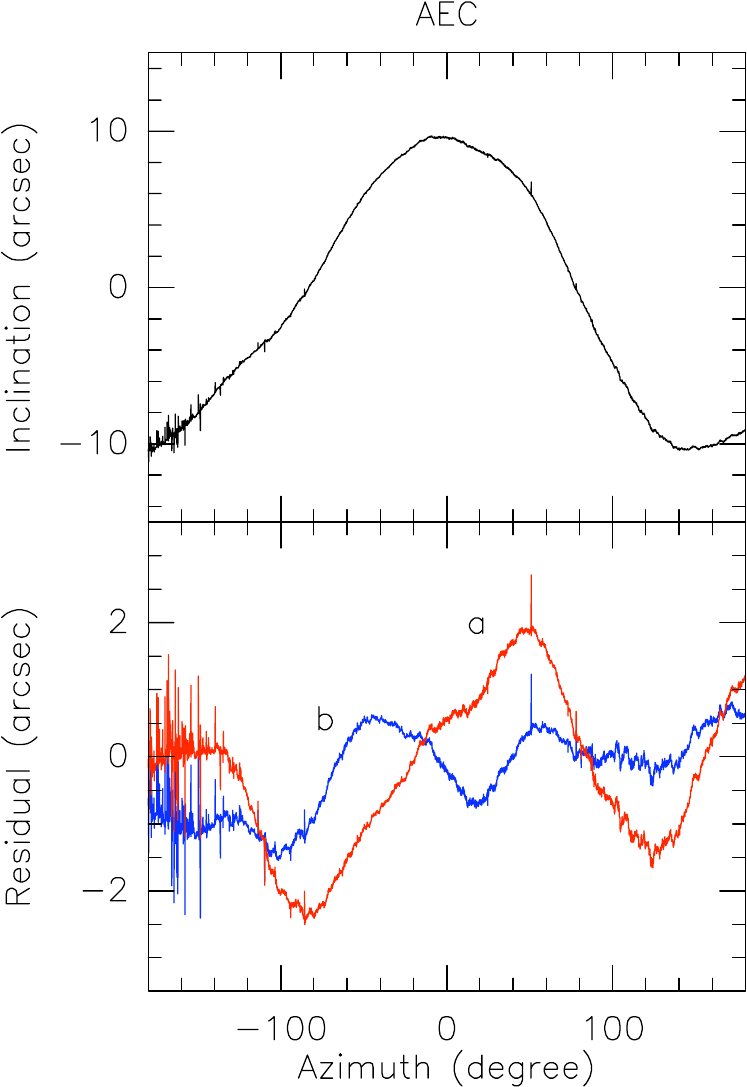}}
\caption{AEC antenna measurements of --\,180$^\circ$ to +\,180$^\circ$ 
AZ rotation (top panel). The curve \textit{a} (lower panel) are the residuals
  of the best--fit a$_{0}$ + a$_{1}$\,sin(A + b$_{1}$), the curve
  \textit{b} are the residuals of the best--fit a$_{0}$ +
  a$_{1}$\,sin(A + b$_{1}$) + a$_{2}$\,sin(2\,A + b$_{2}$).}
\label{fig:aec-azwobble}
\end{figure}

The two--angular response [sin(2\,A)] on the AEC antenna, shown in
Fig.~\ref{fig:aec-azwobble} is attributed to an uneveness of the AZ
bearing. This two--angular wobble amounts to $\pm$\,1\,arcsec. After
elimination of the components sin(A) and sin(2\,A), the residual
repeatable deviation is 0.7\,arcsec (rms). The best--fit values to the
measurement to a$_{0}$ + a$_{1}$\,sin(A + b$_{1}$) + a$_{2}$\,sin(2\,A
+ b$_{2}$) are given in Table \ref{tab:aec-azwobble}.

\begin{table}
\centering
\caption{AEC Antenna: Parameters of the AZ axis Inclination and Wobble}
\begin{tabular}{|cr|cr|}
\hline 
 Linear &    response      &  Linear  & + 2\,A response \\
\hline
 a$_{0}$       & --\,2.65$''$        & a$_{0}$  & --\,2.66$''$    \\
 a$_{1}$       & --\,10.71$''$       & a$_{1}$  & --\,10.75$''$ \\
 b$_{1}$       & --\,81.21$^{\rm o}$ & b$_{1}$  & --\,81.47$^{\rm o}$     \\
               &                     & a$_{2}$  &    0.98$''$  \\
               &                     & b$_{2}$  & 39.13$^{\rm o}$  \\
\hline
 rms(residual) & 0.99$''$       & rms(residual) & 0.73$''$ \\
\hline
\end{tabular}
\label{tab:aec-azwobble}
\end{table}

The installation of an inclinometer on the
center of the azimuth bearing allows an easy, and repeatable
determination of the run-out or wobble of the azimuth bearing, which
can be used in the pointing model as constant terms. This has been
used on many telescopes (\cf\ \cite{Gawronski2000}).  The main
advatage obtained with such an inclinometer on the azimuth bearing
is that it allows an update of the pointing constants describing the
tilt of the AZ axis whenever the slew angle between sources is larger
than $\sim 60$ degrees (\cite{Penalver2001}).

\section{Conclusions}

The measurements indicate that the path length specifications are fulfilled on
both antennas, at least during time intervals of 1/2 to 1 hour. The path 
length variation is primarily due to unavoidable residual thermal
dilatation of the (insulated) antenna steel 
components, and may span $\sim$\,200\,$\mu$m within a day. The path length 
variation can be predicted with high precision from temperature measurements 
at a few positions of the steel components, either used in empirical relations
or the finite element model. It is expected that identical antennas will
experience similar temperature variations of the ambient air so that the
differential effect may even be smaller than stated here. Wind at speeds below
the specification limit (9\,m/s), and OTF and FSW motions of the antennas, do
not degrade the path length stability.

As far as possible to measure, the antennas show similar behaviour of
damping of the thermal environment, \ie\ the ambient air temperature
and the solar radiation. The BUS of the VertexRSI antenna shows a good
temperature homogeneity, even under full exposure to Sun shine. 

Altough the AZ bearings have a higher order azimuth dependent wobble, the
effect can be considered in the pointing model with an accuracy better than
0.6 arcsecond. On the VertexRSI prototype antenna, the wobble was very stable 
with time.

\section*{Acknowledgments}
The data were collected at the VLA site by A. 
Ot\'arola (ESO \& NRAO), N.~Emerson and J.~Cheng (both at NRAO, Tucson, USA). 
The measurement of the antenna when tracking the Sun was made by the Antenna 
Group (NRAO, Tucson, USA). The FEM data of the VertexRSI antenna were provided
by Vertex--Germany (Mr.~Omlor, November 2003) and are based on the original 
ASKA model. The FEM data of the AEC antenna were provided by Mr.~Koch (ESO; 
April 2003) and are based on the model provided by EIE in March 2003. The 
comparison of temperature induced path length variations and FEM predictions 
are based on the FEM models of the antenna steel parts, provided by 
NRAO--Tucson and EIE--ESO, and used by M.~Bremer (IRAM, France). Many data 
were extracted from the DataBase (F.~Staufer, NRAO, Socorro,
USA).


\bibliographystyle{IEEEtran}
\bibliography{IEEEabrv,mybibfile}
%

%

\begin{biography}[{\includegraphics[width=1in,height=1.25in,clip,keepaspectratio]{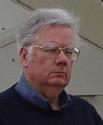}}]{Albert Greve}
was born in Germany.  He studied astronomy at Leiden University,
Leiden, The Netherlands and received the Ph.D. degree from Utrecht
University, Utrecht, The Netherlands, in 1978.

He worked at Culhan Laboratory, UK, the Queens University, Belfast,
Ireland, the Max Planck Institute for Radioastronomy, Germany, and at
Institut de Radio Astronomie Millim\'etrique (IRAM), France and
Spain.  He was involved in the construction and testing of several
radio telescopes, in particular surface adjustments and radio optics,
but also in the thermal design of modern millimeter wavelength
telescopes (and several optical telescopes).  He worked in the field
of radio astronomy, in particular mm-wavelength VLBI.  He retired at
the end of 2003.

Institut de Radio Astronomie Millim\'etrique, 300 rue de la Piscine,
Domaine Universitaire, 38406 Saint Martin d'H\'eres, France
(greve@iram.fr)
\end{biography}

\begin{biography}[{\includegraphics[width=1in,height=1.25in,clip,keepaspectratio]{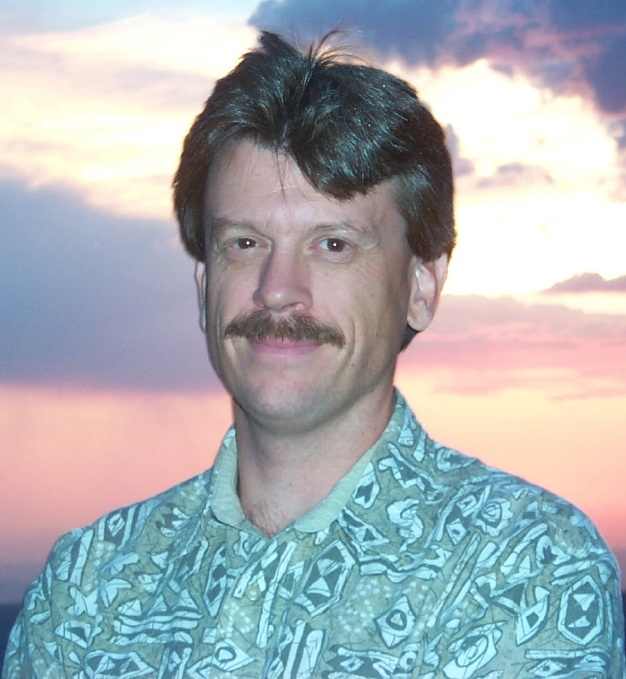}}]{Jeffrey G.~Mangum}
received the Ph.D. degree in astronomy from the University of Virginia
in 1990.  

Following a two-year residency as a postdoctoral researcher in the
astronomy department at the University of Texas he joined the staff of
the Submillimeter Telescope Observatory (SMTO) at the University of
Arizona.  In 1995 he joined the scientific staff at the National Radio
Astronomy Observatory (NRAO) in Tucson, Arizona, and subsequently moved to
the NRAO headquarters in Charlottesville, Virginia.  His research
interests include the astrophysics of star formation, the solar
system, and external galaxies, the performance characterization of
reflector antennas, and calibration of millimeter-wavelength
astronomical measurements.

National Radio Astronomy Observatory, 520
Edgemont Road, Charlottesville, VA  22903, USA (jmangum@nrao.edu)
\end{biography}





\end{document}